\documentclass[twocolumn]{pasj01}
\usepackage[varg]{txfonts}
\usepackage[T1]{fontenc}
\usepackage{comment}
\usepackage{natbib,bm}
\usepackage[switch,mathlines]{lineno}
\begin{document}
\title{Isotopic variation of non-carbonaceous meteorites caused by dust leakage across the Jovian gap in the solar nebula}

\author{Kazuaki A. \textsc{Homma}\altaffilmark{1}, Satoshi \textsc{Okuzumi}\altaffilmark{1}, Sota \textsc{Arakawa}\altaffilmark{2}, and Ryota \textsc{Fukai}\altaffilmark{3}}

\altaffiltext{1}{Department of Earth and Planetary Sciences, 
Tokyo Institute of Technology, Meguro, Tokyo 152-8551, Japan}
\altaffiltext{2}{Japan Agency for Marine-Earth Science and Technology,
3173-25 Showa-machi, Kanazawa-ku, Yokohama, 236-0001, Japan}

\altaffiltext{3}{Institute of Space and Astronautical Science, Japan Aerospace Exploration Agency, Kanagawa 252-5210, Japan}
\email{kzh1226uni@gmail.com, okuzumi@eps.sci.titech.ac.jp}

\maketitle
\begin{abstract}
High-precision isotopic measurements of meteorites revealed that they are classified into non-carbonaceous (NC) and carbonaceous (CC) meteorites.
One plausible scenario for achieving this grouping is the early formation of Jupiter because massive planets can create gaps that suppress the mixing of dust across the gap in protoplanetary disks. 
However, the efficiency of this suppression by the gaps depends on dust size and the strength of turbulent diffusion, allowing some fraction of the dust particles to leak across the Jovian gap. 
In this study, we investigate how isotopic ratios of NC and CC meteorites are varied by the dust leaking across the Jovian gap in the solar nebula.
To do this, we constructed a model to simulate the evolution of the dust size distribution and the $^{54}$Cr-isotopic anomaly $\varepsilon^{54}$Cr in isotopically heterogeneous disks with Jupiter.
Assuming that the parent bodies of NC and CC meteorites are formed in two dust-concentrated locations inside and outside Jupiter's orbit, referred to as the NC reservoir and CC reservoir, we derive the temporal variation of $\varepsilon^{54}$Cr at the NC and CC reservoir.
Our results indicate that substantial contamination of CC materials occurs at the NC reservoir in the fiducial run.
Nevertheless, the values of $\varepsilon^{54}$Cr at the NC reservoir and the CC reservoir in the run are still consistent with those of NC and CC meteorites formed around 2 Myrs after the formation of calcium-aluminum-rich inclusions.
Moreover, this dust leakage causes a positive correlation between the $\varepsilon^{54}$Cr value of NC meteorites and the accretion ages of their parent bodies.
\end{abstract}

\section{Introduction}\label{s:intro}

Planet formation begins with dust evolution in protoplanetary disks.
It is important to reveal how dust grains and aggregates evolve into planets in the solar system.

Meteorites are clues to understanding the dust evolution in the solar system because meteorites retain information about aggregates in the primordial solar nebula.
Recent analyses reveal that meteorites can be classified as non-carbonaceous (NC) and carbonaceous (CC) meteorites by oxygen and nucleosynthetic isotope anomalies (e.g., Ti, Cr, Mo)\footnote{Recently, \cite{Onyett+23} suggest that a silicon isotopic dichotomy is present between achondrites and chondrites. } \citep[e.g.,][]{Warren11,Kleine+20}.

The dichotomy between NC and CC meteorites indicates that the heterogeneous distribution of these isotopes was kept in the primordial solar nebula.
However, the initial heterogeneous distribution would be homogenized by the transportation of aggregates via radial drift and turbulent diffusion \citep{Pignatale+19,Fukai&Arakawa21}.
It is important to reveal how the mixing of the aggregates that formed NC and CC parent bodies is suppressed in the solar nebula.

Many scenarios have been proposed that can explain the two isolated dust reservoirs, including the ones invoking early-formed Jupiter (EJ) \citep[e.g.,][]{Kruijer+2017,Nanne+19,Hutchison+22}, a long-lived pressure bump \citep{Brasser&Mojzsis20,Charnoz+21},  planetesimal formation at two distinct radial locations \citep{Lichtenberg+22, Morbidelli+2022}, and viscous expansion of the nebula \citep{Liu+22,Colmenares+24}. 
The EJ scenario assumes that the solar-system dust reservoir was divided by proto-Jupiter.
It is known that a Neptune-sized or larger protoplanet can open a deep gap in the parent protoplanetary disk \citep[e.g.,][]{Kanagawa+15,Bitsch+18}.
Dust aggregates drifting inward from farther out pile up at the outer edge of the gas gap where the radial gas pressure profile has a local maximum \citep{Desch+18, Pinilla+21}.
Thus, a massive planet can spatially divide a disk into two regions, inside and outside of the planet's orbit.
Gap formation and the dust accumulation can also be an origin of the ring--gap substructures seen in the high-resolution radio and near-infrared images of protoplanetary disks \citep{Bae+23}.
Thermal models of meteorite's parent bodies suggest that NC and CC iron parent bodies formed at $<$ 0.4 Myr and $<$ 1 Myr after calcium-aluminum-rich inclusions (CAI) formation, respectively \citep{Kruijer+2017}.
In the EJ scenario, this constrains the timing of proto-Jupiter formation to less than 1 Myr after CAI formation \citep{Kruijer+2017}.
The EJ scenario has attracted much attention because this scenario can also potentially explain other key features of the solar system.
For example, the EJ could explain the low content of water in the Earth simultaneously because the EJ could have prevented icy aggregates from migrating the current Earth's orbit \footnote{ We note that other scenarios could also explain the low water content of the Earth. For example, Earth-sized planets formed by pebble accretion could be depleted in water if their envelopes are hot enough to sublimate water ice  \citep{Johansen+21,Wang+23}.}
\citep{Morbidelli+16}.
The EJ scenario could also naturally explain the concentration of CAI in CC meteorites \citep{Desch+18}.

However, it is still unknown whether the proto-Jupiter perfectly divided the solar nebula. 
The efficiency of dust accumulation by a planet-induced gap depends on aggregate size and turbulence strength \citep{Zhu+12, Weber+18}.
Turbulence can transport sufficiently small dust particles across the gap.
Thus, some fraction of the particles in the outer dust reservoir for CC meteorites could have been transported to the inner reservoir for NC meteorites.
Previous studies for the EJ scenario \citep{Fukai&Arakawa21, Hutchison+22} demonstrated this effect but without considering dust coagulation and fragmentation.
Recently, \cite{Stammler+23} explored the effect of dust leaking across a gas gap from simulations including dust coagulation, fragmentation, and radial transport in the solar nebula with proto-Jupiter.
They showed a substantial amount of the CC material leaking into the inner disk region by turbulent diffusion and fragmentation.
However, they did not calculate how the leaking affects the isotopic compositions of the two reservoirs. 
Thus, to further test the EJ scenario, one needs to quantify how the radial distribution, mass distribution, and isotopic compositions of the dust in the two reservoirs evolve simultaneously in the presence of the proto-Jupiter.

\begin{figure}[t]
\begin{center}
\includegraphics[width=\hsize, bb = 0 0 720 720]{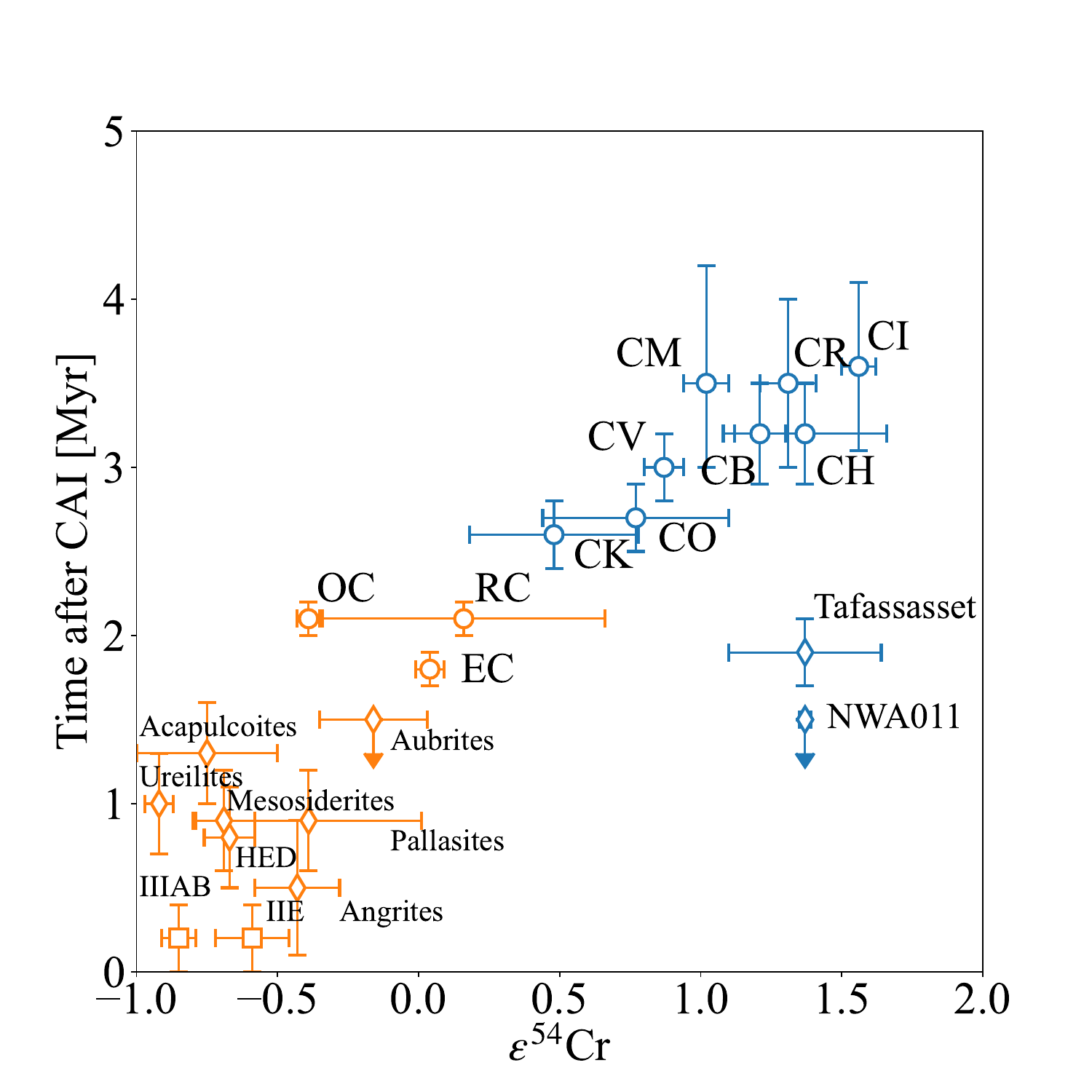}
\end{center}
\caption{The correlation between the $\varepsilon^{54}$Cr values and the accretion ages of meteorites.
The orange and blue plots are the meteorites classified into NC and CC meteorites, respectively. 
The circle, diamond, and square symbols represent chondrites, achondrites, and iron meteorites, respectively.
The values of $\varepsilon^{54}$Cr and the accretion ages for meteorites are listed in table \ref{t:data}, along with the corresponding references. 
The error bars of $\varepsilon^{54}$Cr indicate 95\% confidence intervals.}
\label{f:54Crtime_data}
\end{figure}

In fact, some level of dust leaking across proto-Jupiter's gap is favorable to explain the temporal variation of isotope compositions of meteorites. 
Figure \ref{f:54Crtime_data} shows the correlation between the nucleosynthetic chromium isotope anomalies $\varepsilon^{54}$Cr (see equation (\ref{e:54Cr}) for its definition) of meteorites and the modeled accretion ages of their parent bodies (references in table \ref{t:data}).
The chromium anomaly $\varepsilon^{54}$Cr increases with accretion age, except for NWA 011 and Tafassasset (\citealt{Sugiura&Fujiya14}, reference therein).
Because the CC meteorites have higher $\varepsilon^{54}$Cr, the above temporal trend could imply that some amount of the dust in the CC reservoir was contaminated into the NC reservoir.

The purpose of this work is to investigate whether the EJ scenario can produce the spatial dichotomy and the temporal isotopic variation if we consider dust size evolution with an isotopic anomaly.
To do so, we construct the model to simulate global dust size evolution with chromium isotopic ratio.
We simulate these evolution in the gas disks which have a gap created by a giant planet.
We assess the time evolution of the chromium isotopic ratio at the two dust reservoirs inside and outside the orbit of the planet-induced gas gap, respectively.
Finally, we discuss the effect of the dust leakage on the spatial heterogeneity of $^{54}$Cr in the EJ scenario.

The structure of this paper is as follows.
Our model will be described in section \ref{s:methods}.
The results of dust size evolution with $\varepsilon^{54}$Cr will be presented in section \ref{s:results}.
We will assess the variation of $\varepsilon^{54}$Cr for the NC and CC meteorites in the EJ scenario in this section. 
In section \ref{s:dis}, we will discuss whether the EJ scenario can explain the isotopic dichotomy of the solar system and the correlation between ages and $\varepsilon^{54}$Cr.  
In this section, we will also address improvements of our model and the validity of the adopted parameters.
A summary is provided in section \ref{s:summary}.

\section{Methods}\label{s:methods}
In this section, we describe our model to simulate the radial-1D global dust evolution with coagulation and fragmentation. 
Based on the EJ scenario, we consider a protoplanetary disk with a gap opened by a single giant planet.
We choose $^{54}$Cr as a tracer of isotopic heterogeneity in the solar nebula and simulate the evolution of the $^{54}$Cr abundances of aggregates with different masses and radial distances by taking into account their radial drift, turbulent diffusion, and mutual collisions (coagulation and fragmentation).  
We use the simulation results to assess the evolution of the $^{54}$Cr abundances inside and outside the planet-induced gap as a function of time.

\subsection{Gas disk} \label{s:gas}
We simulate dust evolution in a steady gas disk around a Solar-type star with a single giant planet. Assuming hydrostatic equilibrium, the gas density $\rho_{\rm g}$ is given by $\rho_{\rm g} = \Sigma_{\rm g}/(\sqrt{2 \pi} h_{\rm g})\exp(-z^2/(2h^2_{\rm g}))$, where $\Sigma_{\rm g}$,  $z$ and $h_{\rm g}$ are the gas surface density, the vertical height from the disk midplane and the gas scale height, respectively.
The gas scale height $h_{\rm g}$ is given by $h_{\rm g} = c_{\rm s}/\Omega_{\rm K}$, where $c_{\rm s} $ is the gas sound speed and  $\Omega_{\rm K}$ is the Keplerian frequency.
The sound speed $c_{\rm s}$ is given by $c_{\rm s} = \sqrt{k_{\rm B} T/m_{\rm g}}$ with the Boltzmann constant $k_{\rm B}$ and the mean molecular mass $m_{\rm g} = 2.34$ amu.
We adopt a protoplanetary disk irradiated passively as our disk model.
The gas temperature $T$ at the midplane is given by $T(r) = 163~(r/1 ~\rm au)^{-3/7}$ K \citep{Kusaka+70,Chiang&Goldreich97}. 

We adopt the minimum mass solar nebula model (MMSN) \citep{Hayashi+85} as our background gas disk model. 
The gas surface density $\Sigma_{\rm MMSN}$ is given by $\Sigma_{\rm MMSN} = 1700~(r/1~{\rm au})^{-1.5}~ \rm g~cm^{-2}$, where $r$ is the radial distance from the central star. 

\begin{figure}[t]
\begin{center}
\includegraphics[width=\hsize, bb=0 0 693 641]{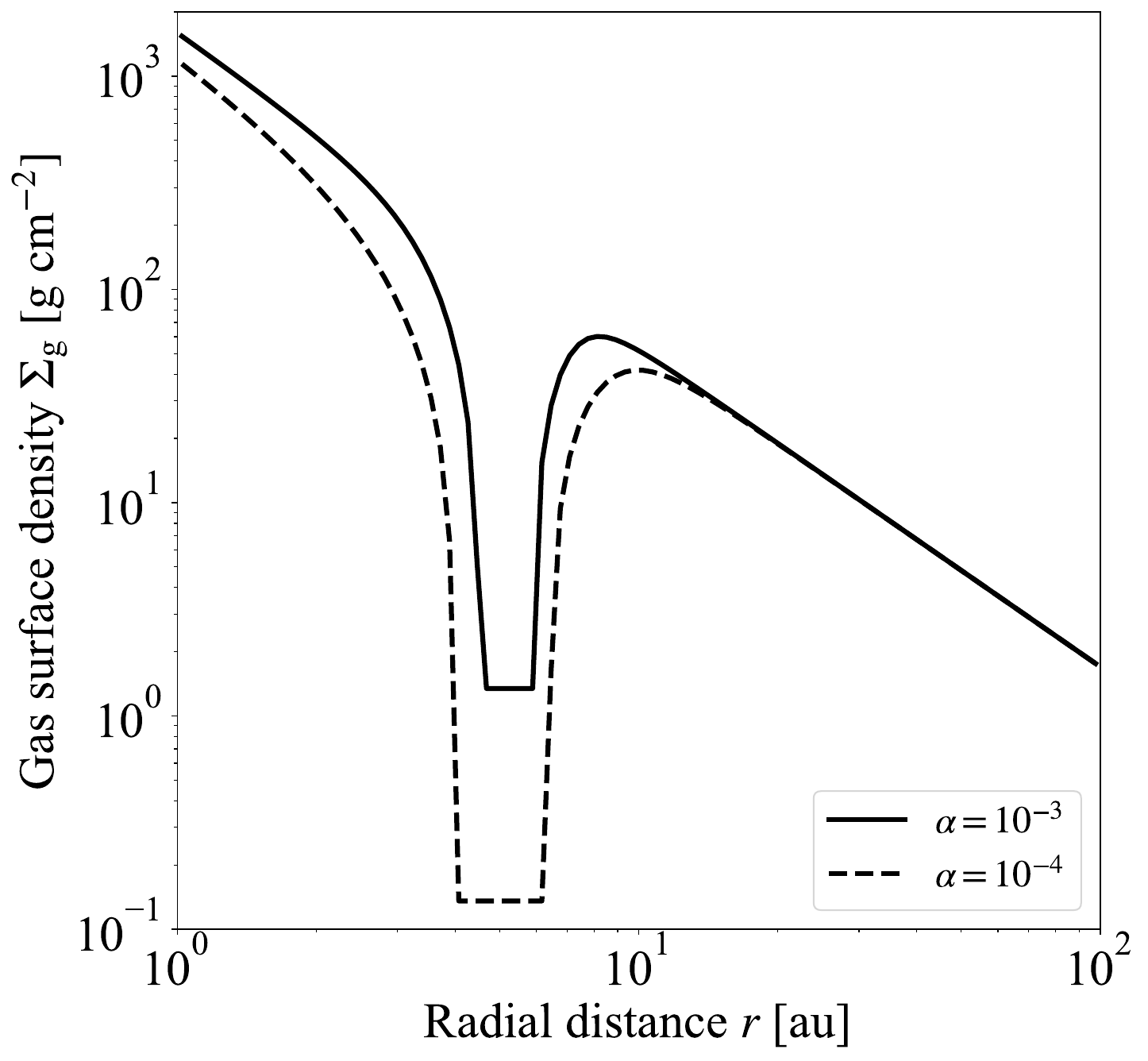}
\end{center}
\caption{Gas surface densities $\Sigma_{\rm g}$ with a gas gap opened by a massive planet at 1 $M_{\rm J}$ and $r = 5.2$ au as a function of $r$ for $\alpha = 10^{-3} $ (solid line) and $10^{-4}$ (dashed line).}
\label{f:sigmag}
\end{figure}

Taking into account the gas gap created by a single giant planet, we apply the gas gap model derived by \cite{Kanagawa+15,Kanagawa+16,Kanagawa+17} to the MMSN model.
\cite{Kanagawa+17} constructed the empirical formula for the gas gap width and the gap depth, respectively (see their equation (6)).
Their formulas are consistent with the 2D hydrodynamic simulation (see figure 2 of \citealt{Kanagawa+17}).
Following \cite{Kanagawa+17}, the gas surface density $\Sigma_{\rm g} $ is given by 
\begin{eqnarray}
\Sigma_{\rm g}(r)= \Sigma_{\rm MMSN} \times\left \{
\begin{array}{l}
\frac{1}{1+0.04K} ~~~ ({\rm for} ~ \left|r-R_{\rm p}\right| < \Delta R) \\
\\
\left\{1+\left(\frac{\Sigma_{\rm gap}}{\Sigma_{\rm MMSN}}\right)^n\right \}^{-1/n}~~{\rm otherwise}
\end{array}
\right., 
\end{eqnarray}
where $R_{\rm p}$ is the orbital radius of the planet, $\Delta R$ is the gap width, and $\Sigma_{\rm gap}$ is the gas surface density at the edge of the gas gap, respectively. The dimensionless parameter $K$
which represents the gap depth is defined by 
\begin{eqnarray}
K  = \alpha^{-1} \left(\frac{M_{\rm p}}{M_{\ast}}\right)^2\left(\frac{h_{\rm p}}{R_{\rm p}}\right)^5,
\end{eqnarray}
where $\alpha$ and $h_{\rm p}$ is a dimensionless parameter that quantifies the strength of turbulence \citep{shakura&Sunyaev73} and the gas scale height at $R_{\rm p}$, respectively.
The gap width $\Delta R$ is given by 
\begin{eqnarray}
\Delta R  = \frac{{K'}^{1/4}R_{\rm p}}{4}\left(\frac{1}{1+0.04K}+0.08\right),
\end{eqnarray}
where the dimensionless parameter $K'$ which is defined by  
\begin{eqnarray}
K'  = \alpha^{-1} \left(\frac{M_{\rm p}}{M_{\ast}}\right)^2\left(\frac{h_{\rm p}}{R_{\rm p}}\right)^3.
\end{eqnarray}
The ratio of  $\Sigma_{\rm gap}$ to $\Sigma_{\rm MMSN}$ is represented as
\begin{eqnarray}
\frac{\Sigma_{\rm gap}}{\Sigma_{\rm MMSN}} = 4{K'}^{-1/4}\frac{\left|r-R_{\rm p}\right|}{R_{\rm p}} - 0.32.
\end{eqnarray}
Our formula for the gas surface density distribution at the gap edge differs from the origin one constructed by \cite{Kanagawa+17}. 
If we choose the origin one, the differential coefficients is discontinuous and diverges to infinity at the edges of the gap.
This may cause the computation error to the calculation.
Thus, we represent $\Sigma_{\rm g}$ with a function that can connect $\Sigma_{\rm MMSM}$ and $\Sigma_{\rm gap}$ smoothly by taking a finite number $n$. 
In this work, we adopt $n = 3$. 
Figure \ref{f:sigmag} shows our gas surface densities at $R_{\rm p}$ = 5.2 au and $M_{\rm p}$ = 1 $M_{\rm J}$ for different $\alpha$ parameters.

We simulate the dust evolution in the existence of a gas giant. 
Thus, our simulation assumes that a gas giant was already formed by the disk gravitational instability during the formation of the star--disk system \citep{Inutsuka+10}. 
For simplicity, we set current Jupiter mass and Jupiter orbit as $M_{\rm p}$ and $R_{\rm p}$ and do not consider the growth and migration of the giant planet.
Gas giants create deep gas gaps \citep{Kanagawa+16,Kanagawa+17}.
Aggregates leak less from deep gas gaps than from the shallow gas gaps \citep{Stammler+23}.
Thus, we simulate the worst leaky case. 
If planets are smaller, the contamination timescale of CC material to NC meteorites would be faster than our results.
\subsection{Dust evolution} \label{s:dustevo}
\subsubsection{Evolutionary equation}
We simulate the evolution of the full dust size distribution in the modeled gas disks by taking into account the radial drift, turbulent diffusion, coagulation, and fragmentation of dust \citep[e.g.,][]{Birnstiel+10,Okuzumi+12,Homma+19}. 
We solve the vertically integrated radial coagulation--advection--diffusion equation with using the bin scheme method developed by \cite{Brauer+08}.
The evolution equation is given by
\begin{eqnarray}
     \frac{\partial \mathcal{N}(r,m)}{\partial t} = & - \frac{1}{r}\frac{\partial}{\partial r}\left[v_{r} \mathcal{N}(r,m) -D_{\rm d}\Sigma_{\rm g}\frac{\partial}{\partial r}\left(\frac{\mathcal{N}(r,m)}{\Sigma_{\rm g}} \right)\right] \nonumber  \\ 
     & + \left. \frac{\partial \mathcal{N}(r,m)}{\partial t}\right|_{\rm coll} \label{e:sizedis},
\end{eqnarray}
where $\mathcal{N}(r,m)$ is the size distribution function defined as the dust surface density per unit mass at aggregate mass $m$ and $r$, $v_{r}$ is the dust radial drift velocity, and $D_{\rm d}$ is the dust turbulent diffusion coefficient. 
The first and second term in the right-hand side of equation (\ref{e:sizedis}) relates to advection of dust via radial drift and to turbulent diffusion of dust, respectively.
The third term of equation (\ref{e:sizedis}) relates to dust coagulation and fragmentation.
The details of the third term are described in section \ref{s:coagfrag}.

\subsubsection{Radial drift and diffusion}\label{s:rdad}
Dust can either lose or gain angular momentum through interactions with the rotating gas disk, leading to inward or outward drift, respectively \citep{Adachi+76,Weidenschilling77}.
The radial drift velocity $v_{\rm r}$ is given by
\begin{eqnarray}
    v_{\rm r} = - \frac{2 \rm St}{1+\rm {St}^2}\eta v_{\rm K}, \label{e:vr}
\end{eqnarray}
where $\eta$ is the dimensionless parameter which characterizes the ratio of the pressure gradient force to the stellar gravity and St is the dimensionless stopping time.
The dimensionless parameter $\eta$ is defined by
\begin{eqnarray}
   \eta \equiv - \frac{1}{2}\left(\frac{c_{\rm s}}{v_{\rm K}}\right)^2\frac{d \ln P}{d \ln r},
\end{eqnarray}
where $v_{\rm K}$ is the Keplerian velocity and $P = \rho_{\rm g} c^2_{\rm s}$ is the gas pressure.
The above definition shows that the drift velocity is proportional to the pressure gradient.
The pressure gradient is positive at the outer edge of gas gaps and negative at the smoothed gas disks. 
Thus, aggregates pile up at the outer edge of the gaps.
We do not consider the radial velocity of the gas into $v_{\rm r}$.
As discussed in section \ref{s:vg}, this treatment would not have a significant impact on our conclusion.

Gas turbulence can diffuse dust radially and affect the efficiency of dust filtration at the gas gaps.
The dust diffusion coefficient $D_{\rm d}$ is written as \citep{Youdin&Lithwick07}
\begin{eqnarray}
    D_{\rm d}  = \frac{\alpha c_{\rm s}h_{\rm g}}{1 + {\rm St^2}}. \label{e:Dd}
\end{eqnarray}

The dimensionless stopping time St which controls the strength of the radial velocity and turbulent diffusion is written as \citep{Okuzumi+12}
\begin{eqnarray}
        {\rm St} = \frac{\rho_{\rm int} a}{\rho_{\rm g}v_{\rm th}}\left(1+\frac{4a}{9\lambda_{\rm mfp}}\right)\Omega_{\rm K},
\end{eqnarray}
where $\rho_{\rm int}$ is the dust internal density, $a$ is the dust radius, $v_{\rm th} = \sqrt{8k_{\rm B} T /\pi m_{\rm g}}$ is the thermal velocity, and $\lambda_{\rm mfp} = m_{\rm g}/(\sigma_{\rm mol}\rho_{\rm g}) $ is the mean free path of gas molecules with the molecular collisional cross-section $\sigma_{\rm mol} = 2 \times 10^{-15} ~\rm cm^2$. 

\subsubsection{Coagulation and fragmentation}\label{s:coagfrag}
Dust monomers grow into aggregates via collisions.
We set the size of a monomer and initial dust-to-gas ratio $\Sigma_{\rm d}/\Sigma_{\rm g}$ to be 0.1 $\micron$ and 0.01, respectively. 
We model the dust monomers and their aggregates as an icy mixture of compact spheres whose internal density $\rho_{\rm int}$ is $\rho_{\rm int} = 1.5 \rm ~g~cm^{-2}$.
The monomer's mass $m_{0}$ and the aggregate's mass $m$ are given by $m_{0} \approx 6.3 \times 10^{-15}$ g and $m = (4\pi/3)\rho_{\rm int}a^3$.

The growing dust aggregates could experience fragmentation at high-speed collisions.
We adopt the fragmentation model developed by \cite{Okuzumi&Hirose12} to describe the outcome of the dust collisions \citep{Okuzumi+16,Homma+19}. 
We refer to sections 3 of \cite{Okuzumi&Hirose12} and section 4.4 of \cite{Okuzumi+16} for details of the model.
When the aggregates whose masses are $m$ and $m'$ ($m > m'$) collide with each other at the collision velocity $\Delta v$, the new aggregate's mass is given by $m+s(\Delta v, v_{\rm f})m'$, where $v_{\rm f}$ is the fragmentation threshold velocity and $s(\Delta v, v_{\rm f}) = \min[1, -{\ln(\Delta v/v_{\rm f})/\ln5}]$ is the sticking efficiency.
For simplicity, we neglect the size distribution of the collisional fragments and assume that all remnants of the two aggregates go into the submicron-sized monomers.
We refer to the newly formed aggregates as remnants and the monomers as fragments.
In reality, the fragments obey a size distribution that depends on dust properties \citep{Arakawa+22, Arakawa+23, Hasegawa+23}.
Our simulation maximizes the amount of dust leaking across the gas gaps. 
If we consider the size distribution of the fragments, the mass of large aggregates that cannot pass through the gap is greater than that without accounting for the distribution.
On the other hand, the large aggregates create fragments continuously.
There are both effects of reducing and increasing the small fragments.
We should simulate our model with the size distribution of fragments in the future. 

Dust collisions are induced by  Brownian motion, radial and azimuthal drift, vertical settling, and turbulence. 
We calculate the collision velocity $\Delta v$ including the contribution of these motions with using equations (16) -- (20) of \cite{Okuzumi+12}.
We refer to sections 2.3.2 of \cite{Okuzumi+12} for details on the collision velocity.

The collision term of equation (\ref{e:sizedis}) is given by 
\begin{eqnarray}
m\left. \frac{\partial \mathcal{N}(r,m)}{\partial t}\right|_{\rm coll}
 &=& - m\int_0^\infty dm' K(r,m,m') \mathcal{N} \mathcal{N}'
 \nonumber \\
 && +  \int_0^\infty dm'\int_0^{m'} dm'' K(r,m',m'') \mathcal{N}' \mathcal{N}'' 
  \nonumber \\
 &&\times \left[ m \delta(m'+s m''-m) 
  \right.
 \nonumber \\
 && \left. + (1-s)m''\delta (m_0-m) \right],
 \label{e:coll}
\end{eqnarray}
where $\mathcal{N}' = \mathcal{N}(r,m')$, $\mathcal{N}'' = \mathcal{N}(r,m'')$, 
and $K(r,m',m'')$ is the vertically integrated collision rate coefficient.
The first term in the right-hand side of equation (\ref{e:coll}) represents the loss of $m \mathcal{N}$ by the growth of aggregates with $m$.
The first and second delta functions in equation (\ref{e:coll}) mean the formation of the remnants at $m$ and the formation of the fragments by the collisions induced by a pair of aggregates, respectively.
This treatment can solve the global dust size evolution with coagulation and fragmentation \citep{Okuzumi&Hirose12,Homma+19}.

The collision rate coefficient is written as 
\begin{eqnarray}
        K(r,m',m'') = \frac{ \sigma_{\rm coll} }{2 \pi h'_{\rm d}h''_{\rm d}} \int^{\infty}_{-\infty} \Delta v \exp \left( -\frac{z^2}{2{h^\ast_{\rm d}}^2} \right) dz,
\end{eqnarray}
where $\sigma_{\rm coll} = \pi(a^2 + {a'}^2)$ is the collisional cross-section, $h'_{\rm d}$ and $h''_{\rm d}$ are the scale heights of dust aggregates at $m'$ and $m''$, and $h^\ast_{\rm d} \equiv ({h'}^{-2}_{\rm d}+{h''}^{-2}_{\rm d})^{-1/2}$. 
Assuming the equilibrium between sedimentation and turbulent diffusion of dust in the vertical direction, $h_{\rm d}$ is written as \citep{Dubrulle+95, Youdin&Lithwick07}
\begin{eqnarray}
        h_{\rm d} = h_{\rm g} \left(1+\frac{\rm St}{\alpha}\frac{1+2\rm St}{1 + \rm St}\right)^{-1/2}.
\end{eqnarray}
\subsection{$\varepsilon^{54}$Cr evolution}\label{s:Crevo}
In addition to the mass evolution given by equation (\ref{e:sizedis}), we also solve the evolution of dust isotopic composition through collisions.
We adopt the chromium isotopic ratio $^{54}$Cr/$^{52}$Cr.
The nucleosynthesis of a chromium isotope $^{54}$Cr in stellar environments has been well-studied by previous studies \citep[e.g.,][]{Trinquier+07, Qin+11, Burkhardt+17, Hellmann+23}.
It is thought that the chromium isotopic heterogeneity is caused by the heterogeneous distribution of presolar oxide grains in the solar nebula \citep{Dauphas+10, Nittler+18}.
Moreover, we choose Cr because the chromium abundances are uniformly between the chondrites \citep{Scott&Krot14,Lodders21}. 
It can give the homogeneous chromium abundance in dust to calculate $^{54}$Cr/$^{52}$Cr (see section \ref{s:Cal54Cr}).

\subsubsection{Evolutionary equation}\label{s:Uevol}
To quantify the abundance of $^{54}$Cr in dust, we denote the average number of $^{54}$Cr atoms in each aggregate as $c(r,m)$. 
The average is taken over aggregates at mass $m$ and radius $r$.   
The number of total $^{54}$Cr atoms per disk area can then be written as $U(r,m) \equiv c(r,m) \mathcal{N}(r,m)$. 
Because the concentration of the presolar oxide
grains determines the $^{54}$Cr abundance in meteorites, $c(r,m)$ also can represent the concentration of the presolar grains in a monomer and an aggregate.

We follow the evolution of $c(r,m)$ using the moment approach developed by \cite{Okuzumi+09}.  
This approach treats the collisional evolution of particles characterized by two parameters $m$ and $Q$, where $Q$ can be any extensive variable (aggregate volume, charge, and so on). In principle, the distribution of particles in the two-dimensional parameter space can be described by the two-dimensional distribution function $f_2(m,Q)$. The mean value of $Q$ for fixed $m$ is given by $\bar{Q}(m) = \int Q f_2dQ/f_1$, where $f_1(m) \equiv \int f_2 dQ$ is the mass distribution function. In our case, $Q$ is the number of $\varepsilon^{54}$Cr atoms in each aggregate, with $\bar{Q}$, $f_1$,  $\bar{Q}f_1$ corresponding to $c$, ${\cal N}$, and $U$, respectively.
The central assumption of this approach is that the distribution of $Q$ for given $m$ is narrow, namely, all aggregates of the same mass $m$ have a similar $Q$ value $\approx \bar{Q}(m)$. 
Under this assumption, the equations that describe the evolution of the zeroth and first moments $f_1(m)$ and $\bar{Q}(m)f_1(m)$ can be written in a closed form. This allows us to follow the evolution of $f_1(m)$ and $\bar{Q}(m)$ at a computational cost less than that one would need to solve the fully two-dimensional evolutionary equation for $f_2(m,Q)$. This approach was originally developed by \citet{Okuzumi+09} for simulating the porosity evolution of aggregates and was later applied by \cite{Stammler+17} for treating the collisional evolution of the CO ice abundance in aggregates.

Applying the moment approach described above, the evolutionary equation for $U(r,m)$ can be written as 
\begin{eqnarray}
\frac{\partial U(r,m)}{\partial t} = & - \frac{1}{r}\frac{\partial}{\partial r}\left[v_{r} U(r,m) -D_{\rm d}\Sigma_{\rm g}\frac{\partial}{\partial r}\left(\frac{U(r,m)}{\Sigma_{\rm g}} \right)\right] \nonumber  \\ 
     & + \left. \frac{\partial U(r,m)}{\partial t}\right|_{\rm coll}.\label{e:Uevol}
\end{eqnarray}
The first two terms in the right-hand side of this equation describe the change in $U$ by the radial drift and the turbulent diffusion of dust, respectively. 
The last term describes the change in $U$ by coagulation and fragmentation and is  given by 
\begin{eqnarray}
\left. \frac{\partial U(r,m)}{\partial t}\right|_{\rm coll} = && - c(r,m) \int_0^\infty dm' K(r,m,m') \mathcal{N} \mathcal{N}' \nonumber \\
&& + \int_0^\infty  dm'\int_0^{m'} dm'' K(r,m',m'') \mathcal{N}' \mathcal{N}'' \nonumber \\
&& ~~ \times \left[c_{\rm rem} \delta(m'+s m''-m)  \right. \nonumber \\
&& ~~~~~~ +c_{\rm frag} \delta (m_0-m) \left. \right], \label{e:Ucoll}
\end{eqnarray}
where $c_{\rm rem} $ and $c_{\rm frag} $ are the average numbers of $^{54}$Cr atoms in the remnants and fragments, respectively (see section~\ref{s:coagfrag} for the definition of remnants and fragments in this work).
In general, the remnants and fragments inherit monomers from both collided aggregates.
The values of $c_{\rm rem} $ and $c_{\rm frag}$ should depend on
the mixing of the monomers that would occur during the aggregate collision. 
However, this mixing process has not been studied in detail so far.
Therefore, we simply assume that the mixing is so efficient that the remnants and fragments have an equal $^{54}$Cr mass abundance.
This assumption leads to the following relationship between $c_{\rm rem}$ and $c_{\rm frag}$, 
\begin{eqnarray}
    \frac{c_{\rm rem}}{m'+sm''} = \frac{c_{\rm frag}}{(1-s)m''} \equiv F_{\rm new},
\end{eqnarray}
where $F_{\rm new}$ represents the $^{54}$Cr abundance of the remnants and fragments. 
This quantity is determined 
by the conservation of the total number of $^{54}$Cr atoms during the collision, namely, 
\begin{eqnarray}
F_{\rm new} = \frac{ c(r, m')+c(r, m'')}{m'+m''}.
\end{eqnarray} 
Using $F_{\rm new}$, we rewrite equation (\ref{e:Ucoll}) as 
\begin{eqnarray}
\label{e:Cnew}
\left. \frac{\partial U(r,m)}{\partial t}\right|_{\rm coll}
 &=& - c(r,m) m\int_0^\infty dm' K(r,m,m') \mathcal{N} \mathcal{N}'
 \nonumber \\
 && +  \int_0^\infty dm'\int_0^{m'} dm'' K(r,m',m'') \mathcal{N}' \mathcal{N}'' 
  \nonumber \\
 &&\times F_{\rm new} \left[ m\delta(m'+s m''-m) 
  \right.
 \nonumber \\
 && \left. + (1-s)m''\delta (m_0-m) \right].
\end{eqnarray}

\subsubsection{$\varepsilon^{54}$Cr calculation}\label{s:Cal54Cr}
We translate $c$ in the previous subsection into the $^{54}$Cr isotopic anomaly.
We express the anomaly as 
\begin{eqnarray}
    \varepsilon^{54}{\rm Cr} \equiv \left\{\frac{\left (^{54}{\rm Cr}/^{52}{\rm Cr} \right)}{\left (^{54}{\rm Cr}/^{52}{\rm Cr} \right)_{\rm std}} -  1 \right \} \times 10^4,
\end{eqnarray}
where $(^{54}{\rm Cr}/^{52}{\rm Cr})_{\rm std}$ is the $^{54}{\rm Cr}/^{52}{\rm Cr}$ ratio of the terrestrial standard. 
The number of $^{52}$Cr atoms in an aggregate is given by $Am$, where $A$ is the number of $^{52}$Cr atoms per unit mass.
We assume a homogeneous abundance of $^{52}$Cr in aggregates because the difference in Cr abundance is narrow between the chondrites \citep{Scott&Krot14,Lodders21}.
We choose Cr abundance in CI chondrite as the Cr abundance in aggregates and set $A = 2.5 \times 10^{19}$ \citep{Lodders21}.
The value of $\varepsilon^{54}$Cr is given as
\begin{eqnarray}
    \varepsilon^{54}{\rm Cr} = \left(\frac{c}{m}\frac{1}{A(^{54}{\rm Cr}/^{52}{\rm Cr})_{\rm std}} - 1\right)\times 10^4.  \label{e:54Cr}
\end{eqnarray}

\begin{figure}[t]
\begin{center}
\includegraphics[width=0.8\hsize, bb = 0 0 625 381]{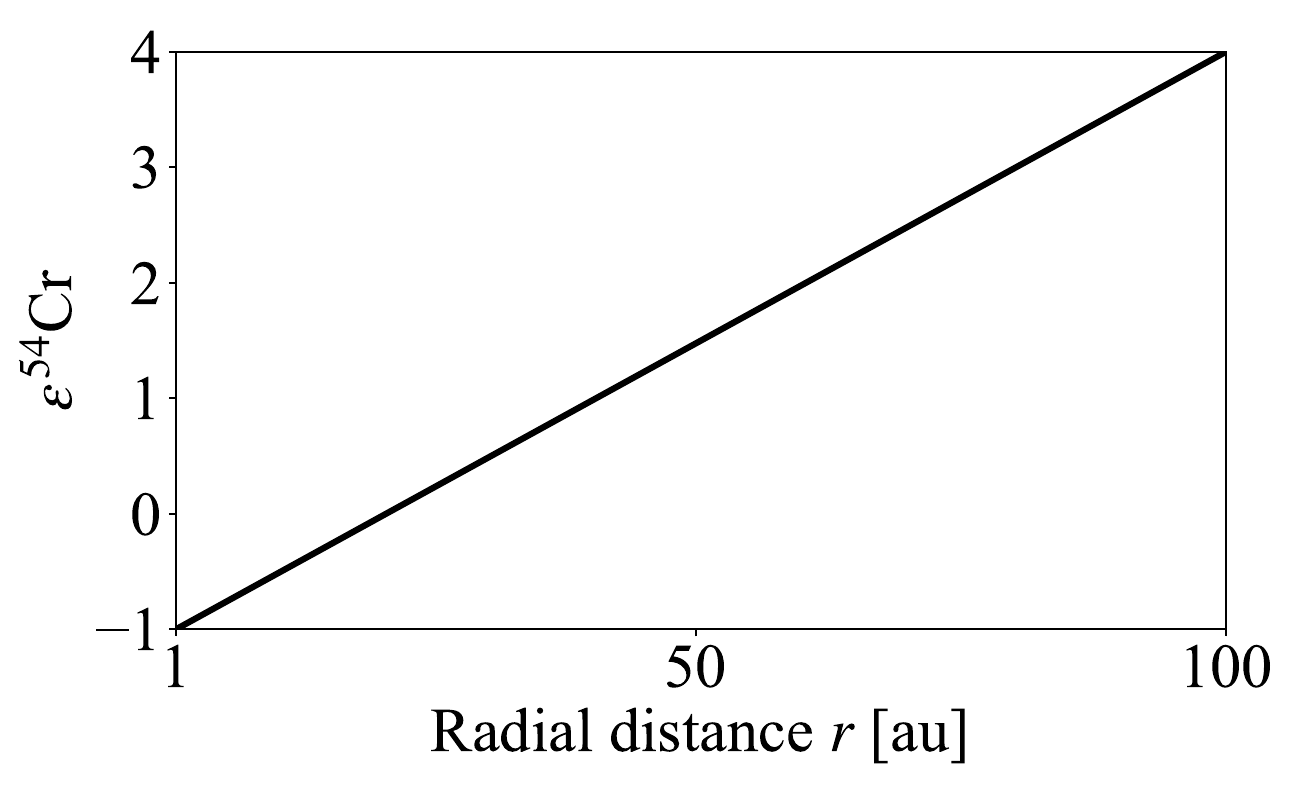}
\caption{Initial radial distribution of  $\varepsilon^{54}$Cr adopted in this study.}
\label{f:54Cr_ini}
\end{center}
\end{figure}

We need to specify the initial distribution of $\varepsilon^{54}$Cr in the solar nebula.
It has been discussed that the origin of heterogeneous isotopic distribution may have been inherited from the isotopic anomaly in the solar parent molecular cloud \citep{Pignatale+19, Nanne+19}, the injection of the presolar grains originated from a supernova into the isotopically homogenous solar disk \citep{Sugiura&Fujiya14, Fukai&Arakawa21}, or the selective thermal destruction of the presolar grains \citep{Trinquier+09, Fukai&Yokoyama17, Ek+20,Liu+22, Colmenares+24}.
However, the exact isotopic value and distribution in the nebula at the beginning of solar system formation are unclear.
Meteoritic data only indicate that the $\varepsilon^{54}$Cr values of the solar nebula solids ranged at least between  $-1$ and 2 (figure \ref{f:54Crtime_data}).
It is possible that the initial range of the $\varepsilon^{54}$Cr values was wider and later narrowed by mixing processes in the nebula.
In this study, we arbitrarily assume that the initial $\varepsilon^{54}$Cr distribution was a linear function of $r$ and ranged from $\varepsilon^{54}{\rm Cr} \approx -1$ at 1 au to $\approx 4$ at 100 au (figure \ref{f:54Cr_ini}).
This assumption specifies the initial radial distribution of $c$ in our simulations.
An initial $\varepsilon^{54}$Cr distribution with a positive radial gradient might be produced by thermal destruction of the presolar neutron-rich carriers by FU-Orionis outbursts \citep{Liu+22,Colmenares+24}.

This initial distribution would not affect the timescale of $\varepsilon^{54}$Cr evolution.
The values of $\varepsilon^{54}$Cr should be affected by the initial condition of $\varepsilon^{54}$Cr.
In our model, the values of $\varepsilon^{54}$Cr at the NC and the CC reservoir are determined by the balance of dust mass and initial $\varepsilon^{54}$Cr values (Detail about these reservoirs in the following section).
If we choose high $\varepsilon^{54}$Cr values at 100 au, the maximum value of $\varepsilon^{54}$Cr would increase at the NC and the CC reservoir.
If we assume a steeper slope of $\varepsilon^{54}$Cr distribution, the maximum value of $\varepsilon^{54}$Cr would decrease at the NC reservoir and increase the CC reservoir.
These are because the values of $\varepsilon^{54}$Cr at the two reservoirs are determined by $\varepsilon^{54}$Cr of dust aggregates which contribute to the reservoirs.  
However, the timescale of $\varepsilon^{54}$Cr evolution is influenced by the dust evolution.
Thus, our results are useful to understand the time variation of $\varepsilon^{54}$Cr by dust evolution.
We also note that the power of $\Sigma_{\rm d}$ is important for $\varepsilon^{54}$Cr because the total dust mass at the two reservoirs is determined by the power of $\Sigma_{\rm d}$.
If disks are smoother than our adopted disks, the maximum values of $\varepsilon^{54}$Cr would decrease at the NC reservoir and the CC reservoir.

\subsection{NC and CC reservoirs} \label{s:eval}
We focus on the two spatial regions where aggregates pile up.
Planetesimals which are the building blocks for the parent bodies of meteorites are formed via the streaming instability \citep{Youdin&Goodman05,Johansen+07,Yang+17,Li+21} and/or the gravitational instability \citep{Goldreich&Ward73,Sekiya83}.
These instabilities are triggered at the locations where aggregates are piled up, which we call dust reservoirs. 
We assume that the parent bodies of NC and CC meteorites were formed at the dust reservoirs inside and outside Jupiter's orbits (NC and CC reservoirs), respectively.
We do not directly compute planetesimal formation but assume that the NC and CC reservoirs trapped aggregates and facilitated planetesimal formation via the above instabilities.

\begin{figure}
\begin{center}  \includegraphics[width=\hsize, bb = 0 0 268 122]{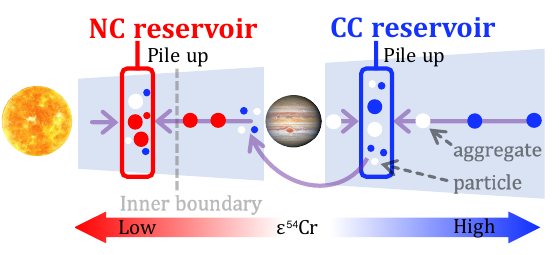}
\end{center}
\caption{Schematic illustration of the NC and CC reservoirs assumed in our model.
    Drifting aggregates are assumed to pile up at two pressure bumps inside and outside Jupiter's orbit, which we call the NC and CC reservoirs, respectively.
    We do not specify the mechanism forming the NC reservoir (see text for a possible candidate) and simply assume that it is located inside the inner boundary of the computational domain (1 au) and traps all dust. The CC reservoir is the pressure bump at the outer edge of the gas gap carved by Jupiter.
    Turbulent diffusion transports small dust particles at the CC reservoir across the Jovian gap.
    For simplicity, aggregates and particles are assumed to be trapped perfectly at the NC reservoir. 
    The possible effect of dust loss on $\epsilon^{54}$Cr at the NC reservoir is discussed in section \ref{s:d_inn}.
    }
    \label{f:sch}
\end{figure}

Figure \ref{f:sch} illustrates the concept of the NC and CC reservoirs in our model, depicting the processes of dust drifting, trapping at the two reservoirs, and leaking from the CC reservoir.
In our model, the CC reservoir is identified as the outer edge of the gas gap formed by Jupiter.
This gap's outer edge naturally serves as a dust reservoir, collecting radially drifting aggregates (see section \ref{s:rdad}).
The location of the NC reservoir is less obvious. The gap's inner edge does not produce a gas pressure maximum and therefore does not trap dust. 
All aggregates passing through the gas gap continue drifting inward until they reach another dust-trap created by a different mechanism.
For example, theory suggests that the magnetorotational instability (MRI; \citealt{BalbusHawley91}) operates in the inner disk region of $T \gtrsim 1000~\rm K$ \citep{Gammie96,DeschTurner15} and produces a dust-trapping pressure bump at the outer boundary of the MRI-active region \citep[e.g.,][]{KretkeLin07,Kretke+09,Ueda+19}.
However, extending our computational domain down to the border of the MRI-active region is computationally expensive because the border would lie at $r \ll 1~{\rm au}$ under our adopted temperature profile, demanding significantly small timesteps (in our simulations, the timestep scales roughly with the local Keplerian timescale; see section \ref{s:set} for details). 
For this reason, we simply assume that there is the NC reservoir interior to the inner boundary of the computational domain (located at $r \approx $ 0.01 au by the adopted model of temperature, see section \ref{s:gas}) and traps all aggregates that have passed the inner boundary.
We also assume that the aggregates get trapped as soon as they pass the inner computational boundary because the timescale of radial drift in the inner region is short. 
The validity of assuming perfect dust trapping at the NC reservoir is discussed in section \ref{s:d_inn}.
It should be noted that we assume the NC parent bodies forming at the border of the MRI-active region as the first step in calculating $\varepsilon^{54}$Cr.
There are many possibilities to reproduce the NC reservoir at other orbits \citep{Schoonenberg&Ormel17,Drkazkowska&Alibert17}. Moreover, the Fe/Ni ratios of NC differentiated meteorites suggest that their parent bodies formed at orbits beyond the water snowline \citep{Grewal+24}.
Thus, the NC bodies might not necessarily form at the border.
However, if a long-lived pressure bump beyond the border can be sustained by some mechanism, results similar to ours might still be achievable. Therefore, calculating the $\varepsilon^{54}$Cr of NC meteorites under this assumption remains useful.

We measure $\varepsilon^{54}$Cr value at the NC and CC reservoirs. The averaged number of $^{54}$Cr atom per dust mass is given by the ratio of the total surface density to the total number of $\varepsilon^{54}$Cr per disk area, $ \int U dm /\int m \mathcal{N} dm$. 
The value of $\varepsilon^{54}$Cr values is given by substituting $ \int U dm /\int m \mathcal{N} dm$ into $c/m$ in equation (\ref{e:54Cr}).

Our results are compared with the actual $\varepsilon^{54}$Cr values and the accretion ages for the parent bodies of meteorites.
Data for the $\varepsilon^{54}$Cr values of meteorites is taken from the compiled data \citep{Burkhardt+17}.
The accretion ages of the parent bodies of chondrites and iron meteorites are constrained by the thermal evolution model of planetesimals \citep{Sugiura&Fujiya14,Kruijer+14}.
The thermal evolution model estimates the maximum temperature in planetesimals heated internally by $^{26}$Al decay.
The peak temperature at which meteorites are experienced is estimated by the degree of metamorphism for chondrites, the chemical composition of iron meteorites, and so pon (see \citealt{Sugiura&Fujiya14, Kruijer+14}).
Thus, the thermal evolution model can constrain the accretion age from the estimated amount of $^{26}$Al at which the meteorite's parent bodies formed.
We use these accretion ages as the formation ages of parent bodies.
Figure \ref{f:54Crtime_data} shows the adapted accretion ages and $\varepsilon^{54}$Cr values for meteorites.
The detailed values and references are listed in table \ref{t:data}.
\begin{table}[ht]
\tbl{Summary of $\varepsilon^{54}$Cr for meteorites and accretion ages for their parent bodies. \footnotemark[$*$] }{
\begin{tabular}{lcc}
   Sample & $\varepsilon^{54}$Cr & Accretion age   \\
   \hline \hline
   Non-Carbonaceous meteorites &  &   \\
   Ordinary chondrite (OC) & $-0.39$ $\pm$ 0.04 & 2.1 $\pm$ 0.1 Myr       \\
   Enstatite chondrite (EC)&  0.04 $\pm$ 0.05 & 1.8 $\pm$ 0.1 Myr       \\
   Rumuruti chondrite (RC) &  0.16 $\pm$  0.5 & 2.1 $\pm$ 0.1 Myr           \\
   Acapulcoites-lodranites & $-0.75$ $\pm$ 0.25 & 1.3 $\pm$ 0.3 Myr            \\
   Ureilites               & $-0.92$ $\pm$ 0.05 & 1.0 $\pm$ 0.3 Myr           \\
   Angrites                & $-0.43$ $\pm$ 0.15 & 0.5 $\pm$ 0.4 Myr\\
   Aubrites                & $-0.16$ $\pm$ 0.19 & < 1.6 Myr\\
   Mesosiderites           & $-0.69$ $\pm$ 0.11 & 0.9 $\pm$ 0.3 Myr\\
   HED                     & $-0.67$ $\pm$ 0.09 & 0.8 $\pm$ 0.3 Myr\\
   Pallasites              & $-0.39$ $\pm$  0.4 & 0.9 $\pm$ 0.3 Myr\\
   IIIAB iron              & $-0.85$ $\pm$ 0.06 & 0.2 $\pm$ 0.2 Myr\\
   IIE   iron              & $-0.54$ $\pm$ 0.13 & 0.2 $\pm$ 0.2 Myr \\
   Carbonaceous meteorites          &  &   \\
   CI          & 1.56 $\pm$ 0.03 & 3.6 $\pm$ 0.5 Myr \\
   CM          & 1.02 $\pm$ 0.08 & 3.5 $+0.7/-0.5$ Myr \\
   CO          & 0.77 $\pm$ 0.33 & 2.7 $\pm$ 0.2 Myr \\
   CV          & 0.87 $\pm$ 0.07 & 3.0 $\pm$ 0.2 Myr \\
   CK          & 0.48 $\pm$ 0.30 & 2.6 $\pm$ 0.2 Myr \\
   CR          & 1.33 $\pm$ 0.10 & 3.5 $\pm$ 0.5 Myr \\
   CH          & 1.37 $\pm$ 0.29 & 3.2 $\pm$ 0.3 Myr \\
   CB          & 1.21 $\pm$ 0.09 & 3.2 $\pm$ 0.3 Myr \\
   Tafassasset & 1.37 $\pm$ 0.27 & 1.9 $\pm$ 0.2 Myr \\
   NWA 011     & 1.37 $\pm$ 0.02 & < 1.6 Myr \\
   \hline
\end{tabular}}\label{t:data}
\begin{tabnote}
\footnotemark[$*$] The values of $\varepsilon^{54}$Cr for all meteorites except Tafassasset and NWA 011 are from the data compiled in \cite{Burkhardt+17} and references therein.  
The data for Tafassasset and NWA 011 are from \cite{Gopel+15} and \cite{Bogdanovski&Lugmair04}, respectively.
Uncertainties of $\varepsilon^{54}$Cr represent 95$\%$ confidence interval estimated with Student's t-distribution or given by two standard errors.
Accretion ages for the parental bodies of all meteorites except iron meteorites are from \cite{Sugiura&Fujiya14}. The ages of IIIAB and IIE iron meteorites are from \cite{Kruijer+14}.
\end{tabnote}
\end{table}

\subsection{Simulation runs}\label{s:param}
\subsubsection{Setting}\label{s:set}
Following \cite{Okuzumi+12}, we adopt an explicit time-integration scheme and a fixed-bin method to solve equations (\ref{e:sizedis}) and (\ref{e:Uevol}).
The computational domain ranges between $r = 1$ and $100$ au and is divided into 100 grid cells with an equal logarithmic width.
The timestep $\Delta t$ is dynamically calculated to ensure that $\mathcal{N}$ and $U$ become non-negative throughout the simulation. 
In our simulation, $\Delta t$ scales with the local Keplerian timescale at the inner boundary of the computational domain because $\Delta t$ is given by the collisional term of these equations.

We impose that the flux is zero 
at the outer boundary and that the diffusive flux is zero at the inner boundary.
The effects of the assumption at the inner boundary on our results are discussed in section \ref{s:d_inn}

\subsubsection{Adopted parameters}
We perform three simulation runs with different values of $\alpha$ and $v_{\rm f}$ to study the effects of turbulence strength and dust stickiness on the isotopic evolution at the NC and CC reservoirs (table \ref{t:param}).
These two parameters determine the efficiency of the dust trapping at the gas gap because they control the level of radial dust diffusion and the maximum aggregate size, the latter of which affects the aggregates' radial drift speed. \citep{Zhu+12}. 
If $\alpha$ is high and/or $v_{\rm f}$ is low, even the largest aggregates would not pile up at the edge of gas gaps. 
In the EJ scenario, dust filtering at the planet-induced gap is essential for creating the isotopic dichotomy.  
In addition, the dust stickiness controls the timescale of the radial redistribution of materials through radial drift. 

Our fiducial run with $\alpha = 10^{-3}$ and $v_{\rm f} = 10~\rm m~s^{-1}$ represents the case where a substantial amount of dust leaks from outside to inside the gas gap. 
The weak turbulence run adopts a lower turbulence strength, corresponding to the situation where the leaking of materials across the gap is suppressed. 
The low stickiness run demonstrates that the aggregates do not pile up at the gas gap with a lower fragmentation velocity.

\begin{table}[t]
\tbl{Parameter values adopted in three simulation runs}{ 
\begin{tabular}{lcc}
   Run name & $\alpha$ & $v_{\rm f} ~\rm (m~s^{-1})$    \\
   \hline \hline
   fiducial  & $10^{-3}$ & 10  \\
   weak turbulence  & $10^{-4}$ & 10 \\
   low stickiness  & $10^{-3}$ & 1 \\
   \hline
\end{tabular}}\label{t:param}
\end{table}

\section{Results}\label{s:results}
\subsection{Fiducial run}\label{s:results_F}
\subsubsection{Evolution of dust mass and $\varepsilon^{54}$Cr distribution}\label{s:results_3d}
\begin{figure*}[ht]
\begin{center}
\includegraphics[width=\hsize,bb = 0 0 2023 504]{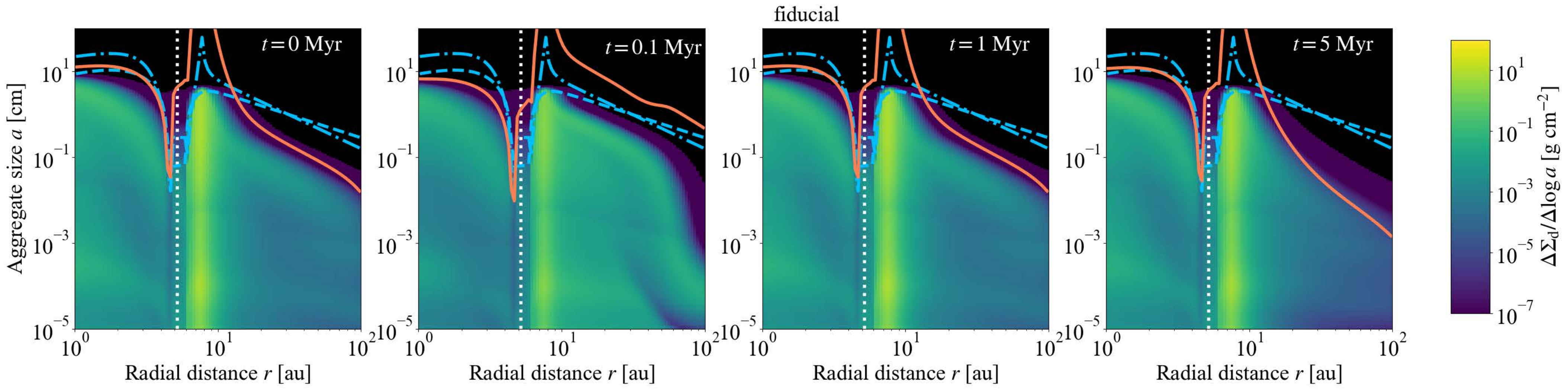}
\end{center}
\begin{center}
\includegraphics[width=\hsize, bb = 0 0 2023 504]{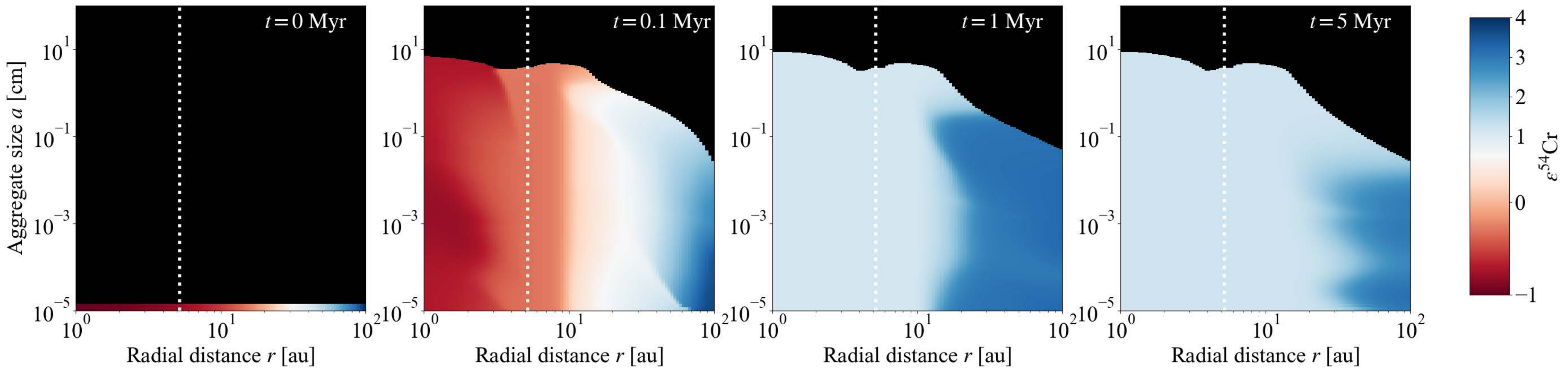}
\end{center}
   \caption{The snapshots of the size distribution $\Delta \Sigma_{\rm d}/\Delta \log m $ (upper panels) and $\varepsilon^{54}$Cr (lower panels) for different times in the fiducial run ($\alpha = 10^{-3}$ and $v_{\rm f} = 10 ~ \rm m~s^{-1}$), respectively. The red line, blue dashed line, and blue dashed-dotted line in the upper panels show the maximum aggregate size set by radial drift, radial drift-induced fragmentation, and turbulence-induced fragmentation, respectively \citep{Birnstiel+12}. The orbit of Jupiter is represented in the white dotted line.}
\label{f:global_vf10a3}
\end{figure*}

Figure \ref{f:global_vf10a3} shows the snapshots of the size distribution (upper panels) and $\varepsilon^{54}$Cr (lower panels) at different times in the fiducial run (table \ref{t:param}), respectively. 
In this work, the size distribution is represented by dust surface density per decade of aggregate radius $\Delta \Sigma_{\rm d}/\Delta \log ~a \equiv m^2 \ln(10)\mathcal{N}$.

In general, aggregates grow until their size reaches the upper limit set by fragmentation or radial drift \citep[e.g.,][]{Brauer+08,Birnstiel+10, Pinilla+21}.
The blue dashed line and blue dashed-dotted line in the upper panels of figure \ref{f:global_vf10a3} indicate the maximum size set by fragmentation when the collision velocity is dominated by radial drift and turbulence.
The maximum Stokes number gives the maximum aggregate size because St is an increasing function of $a$. 
The maximum Stokes number set by turbulence-induced fragmentation $\rm St_{\rm frag,t}$ is given by  \citep{Birnstiel+12} 
\begin{eqnarray}
    {\rm St}_ {\rm frag,t} = \frac{v^2_{\rm f}}{3\alpha c^2_{\rm s}} \approx 0.06 \left(\frac{v_{\rm f}}{10 ~ \rm m~s^{-1}}\right)^2\left(\frac{\alpha}{10^{-3}}\right)^{-1}\left(\frac{r}{1~\rm au}\right)^{3/7}.\label{St_frag_t}
\end{eqnarray}
The maximum Stokes number set by drift-induced fragmentation $\rm St_{\rm frag,r}$ is estimated as \citep{Birnstiel+12}
\begin{eqnarray}
    {\rm St}_ {\rm frag,r} &&= \frac{2v_{\rm f}}{|\eta v_{\rm K}|} \approx 0.3 \left(\frac{v_{\rm f}}{10 ~ \rm m~s^{-1}}\right)\left(\frac{|d\ln P/d\ln r|}{3}\right)^{-1}\left(\frac{r}{1~\rm au}\right)^{-1/14}. \label{St_frag_r}
\end{eqnarray} 
The red line in the upper panels of figure \ref{f:global_vf10a3} indicates the maximum dust size set by the balance between the timescales of radial drift and coagulation. 
The maximum Stokes number set by radial drift is given as \citep{Birnstiel+12} 
\begin{equation}
    {\rm St_{\rm drift}} \approx 0.05 \left(\frac{\epsilon}{10^{-4}}\right)\left(\frac{|d\ln P/d \ln r|}{3}\right)^{-1}\left(\frac{r}{1~{\rm au}}\right)^{-4/7}, \label{St_drift}
\end{equation}
where $\epsilon$ is the vertically integrated dust-to-gas mass ratio.
Initially, aggregates grow to the size set by $\rm{St}_ {frag,t} $ or $ \rm{St}_ {frag,r}$.
The fully grown aggregates then drift inward, causing the dust surface density to decrease except at the outer edge of the planet-induced gas gap.
Thus, the maximum size is determined by $\rm St_{drift}$ at $t \gtrsim 0.1 \rm ~Myr$ at the region where aggregates drift inward.
The upper panels of figure \ref{f:global_vf10a3} show the maximum size is consistent with the size given by the smallest of $\rm {\rm St}_{\rm frag,t}$, $ \rm {\rm St}_ {\rm frag,r},$ and $ {\rm St}_{\rm drift}$.

Figure \ref{f:global_vf10a3} shows that $\Delta \Sigma_{\rm d}/\Delta \log a$ for $t > 0.1$ Myr has a peak at $r \approx 7$ au, where the planet-induced pressure bump traps aggregates drifting in from further out.
Aggregates piling up at the gap outer edge can be more clearly seen in figure \ref{f:sigmad}, where the radial profiles of the size-integrated dust surface density $\Sigma_{\rm d}$ at $t=$ 1 and 5 Myr from the fiducial run are shown by the solid and dashed black lines, respectively. 

However, all dust does not pile up at the pressure bump \citep{Zhu+12,Drkazkowska+19}.
Smaller aggregates have a lower drift velocity toward the pressure maximum at the outer edge (see equation (\ref{e:vr})) and therefore are easier to leak across the gap.
Figure \ref{f:dMddot} shows a snapshot of the radial inward dust mass flux per logarithmic aggregate size, $\Delta \dot{M_{\rm d}}/\Delta \log{a}$, at the outer edge of the gas ($r = 6.9$ au).
The figure shows that aggregates with $\rm St/\alpha < 5$ are allowed to pass through the gap.
Larger aggregates have a vanishingly small radial flux because outward dust drift ($v_{\rm r} > 0$ at 6.9 au) and inward turbulent diffusion cancel out
\footnote{The large $\Delta \dot{M_{\rm d}}/\Delta \log{a}$ at $a \approx$ 1 cm seen in figure \ref{f:dMddot}  does not contradict the $\rm St/\alpha < 5$ criterion mentioned above. While cm-sized aggregates are scarce at $r =$ 6.7 au, they are abundant at $r =$ 7 au (see the black and orange dotted lines in figure \ref{f:dMddot}). This positive radial gradient in $\Delta \Sigma_{d} / \Delta \log a$ produces the inward diffusive flux of the cm-sized aggregates.
However, these aggregates experience collisional fragmentation as soon as they cross $r =$ 6.7 au because their sizes are larger than the size determined by $\rm St_{\rm frag}$ at that location. 
Therefore, these aggregates do not actually pass through the gas gap.}.
For reference, the contribution to $\Delta \dot{M_{\rm d}}/\Delta \log{a}$ from diffusion only is also plotted in figure \ref{f:dMddot}. 
Our result is consistent with the criterion $\rm St/\alpha \gtrsim 1$ for dust filtration by a planet-induced gap to dominate over turbulent diffusion \citep{Zhu+12,Weber+18}.
At the bump, aggregates grow to the size determined by $\rm St_{frag,t} \approx 0.1$, which is much higher than the assumed turbulence parameter $\alpha = 10^{-3}$ (figure \ref{f:global_vf10a3}).
Although the pressure bump traps these largest aggregates, they produce smaller fragments through mutual collisions, and the fragments with ${\rm St}/\alpha < $5 pass through the gap.
This allows a substantial amount of dust to leak into the inner disk region across the gap \citep{Stammler+23}. 

The dust leaking is also observed in the upper panels of figure \ref{f:global_vf10a3}.
They show that there are aggregates larger than $10^{-1}$ cm inside the gas gap at $ t \geq 0.1 $ Myr.
These large aggregates form through the coagulation of small fragments that have leaked through the gap by turbulent diffusion. 
These aggregates continue drifting toward the central star.

The lower panels of figure \ref{f:global_vf10a3} show the evolution of $\varepsilon^{54}$Cr$(r,m)$ in the fiducial run.
It can be seen that $\varepsilon^{54}$Cr of aggregates lying at the same radial position exhibits little size dependence.
As we have mentioned earlier, aggregates first grow to the size set by fragmentation or radial drift and then drift inward.
This means that the timescale of collisions is shorter than that of radial drift.
Thus, local collisions between aggregates are frequent enough to homogenize the $\varepsilon^{54}$Cr values at the same orbit before aggregates drift.
At $t \leq 1$ Myr, $\varepsilon^{54}$Cr at all $r$ increases with time. 
This is simply because of the inward drift of aggregates that initially had a radially positive gradient of $\varepsilon^{54}$Cr. 
At $t \gtrsim$ 1 Myr, in contrast, aggregates lying interior to $r \approx$ 10 au have a $^{54}$Cr abundance that is uniform in both space and time. 
This uniform value is determined by the average $\varepsilon^{54}$Cr value of the small dust particles that leaked through the gap. 

\subsubsection{Evolution of $\varepsilon^{54}$Cr at two reservoirs}\label{s:Crevol_vf10a3}
As mentioned in section \ref{s:eval}, we assume that the CC and NC reservoirs lie at the gap's outer edge and the inner computational boundary, respectively. 
Here, we show how the isotopic compositions of the two reservoirs evolve with time in the fiducial run.
The blue line in figure \ref{f:54Crtime_vf10a3} shows $\varepsilon^{54}$Cr value as a function of time at the CC reservoir. 
Figure \ref{f:54Crtime_vf10a3} indicates that the $\varepsilon^{54}$Cr value at the CC reservoir rapidly increases until $t \approx $ 0.7 Myr and reaches around 0.8 by $t = 5 $ Myr. 
This rapid increase can be attributed to the accumulation of aggregates with a high $^{54}$Cr abundance that drifts from the outer region of the gas disk. 
As a result, the $\varepsilon^{54}$Cr value at the CC reservoir continues to increase over time until most of the aggregates in the gas disk have piled up at the CC reservoir.

The value of the $\varepsilon^{54}$Cr at the CC reservoir does not evolve after the drift timescale of the aggregates exterior to the gas gap (having lower $\varepsilon^{54}$Cr values). 
The drift timescale can be estimated as 
\begin{eqnarray}
    t_{\rm drift} &&= \frac{r}{|v_{\rm r}|} = \frac{1}{2 \rm St} \frac{r}{|\eta| v_{\rm K}} \nonumber \\
    &&\approx 400  \left(\frac{\rm St}{0.2}\right)^{-1}\left(\frac{|d\ln P/d\ln r|}{3}\right)^{-1}\left(\frac{r}{1~\rm au}\right)^{13/14} ~ \rm yr  \label{e:timescale_drift}.
\end{eqnarray} 
In the fiducial run, aggregates initially grow to be the size set by drift-induced fragmentation.
With $\rm St_{frag,r} \approx 0.2$, the drift timescale for the largest aggregates is calculated as $ t_{\rm drift} \approx 0.3$ Myr at the outer computation boundary ($r$ = 100 au). 
This timescale is comparable to the timescale on which $\varepsilon^{54}$Cr of the CC reservoir has increased to the final value.

The orange line in figure \ref{f:54Crtime_vf10a3} shows the $\varepsilon^{54}$Cr values as a function of time at the NC reservoir. 
The $\varepsilon^{54}$Cr value at the NC reservoir initially increases because the NC reservoir acquires the dust particle leaking from the CC reservoir through the gap.
To confirm the contamination of CC materials in this fiducial case, we plot in figure \ref{f:Md_NC} the evolution of total dust mass at the NC reservoir.
The figure shows the mass at the NC reservoir increases with time and reaches $\approx$ 19 $\rm M_{\oplus} $ and $t = 5$ Myr.
This final total mass is larger than the dust mass that was initially contained interior to the gap $\approx 3 \rm M_{\oplus} $. 
This demonstrates that substantial contamination of CC materials occurs at the NC reservoir in this simulation run.

\subsection{Weak turbulence and low stickiness runs} \label{s:results_FaW}
From the results of the fiducial run, it is evident that dust fragmentation and leakage across the planet-induced gap by diffusion control the contamination of CC materials into the NC reservoir.
Here, we study how the simulation results change as we vary turbulence strength and the stickiness of aggregates.

Both turbulence strength and the stickiness directly affect the maximum size of the aggregates.
In the weak turbulence run, aggregates reach the size given by $\rm St_{\rm drift}$ (equation (\ref{St_drift})) because neither their turbulence-induced nor radial drift-induced collision velocity is high enough to cause fragmentation at smaller sizes.
In the low stickiness run, aggregates grow to the size set by $\rm St_{\rm frag,t}$ (equation (\ref{St_frag_t})) due to their low stickiness. 

\begin{figure*}[ht]
\begin{center}
\includegraphics[width=0.4\hsize,bb = 0 0 691 654]{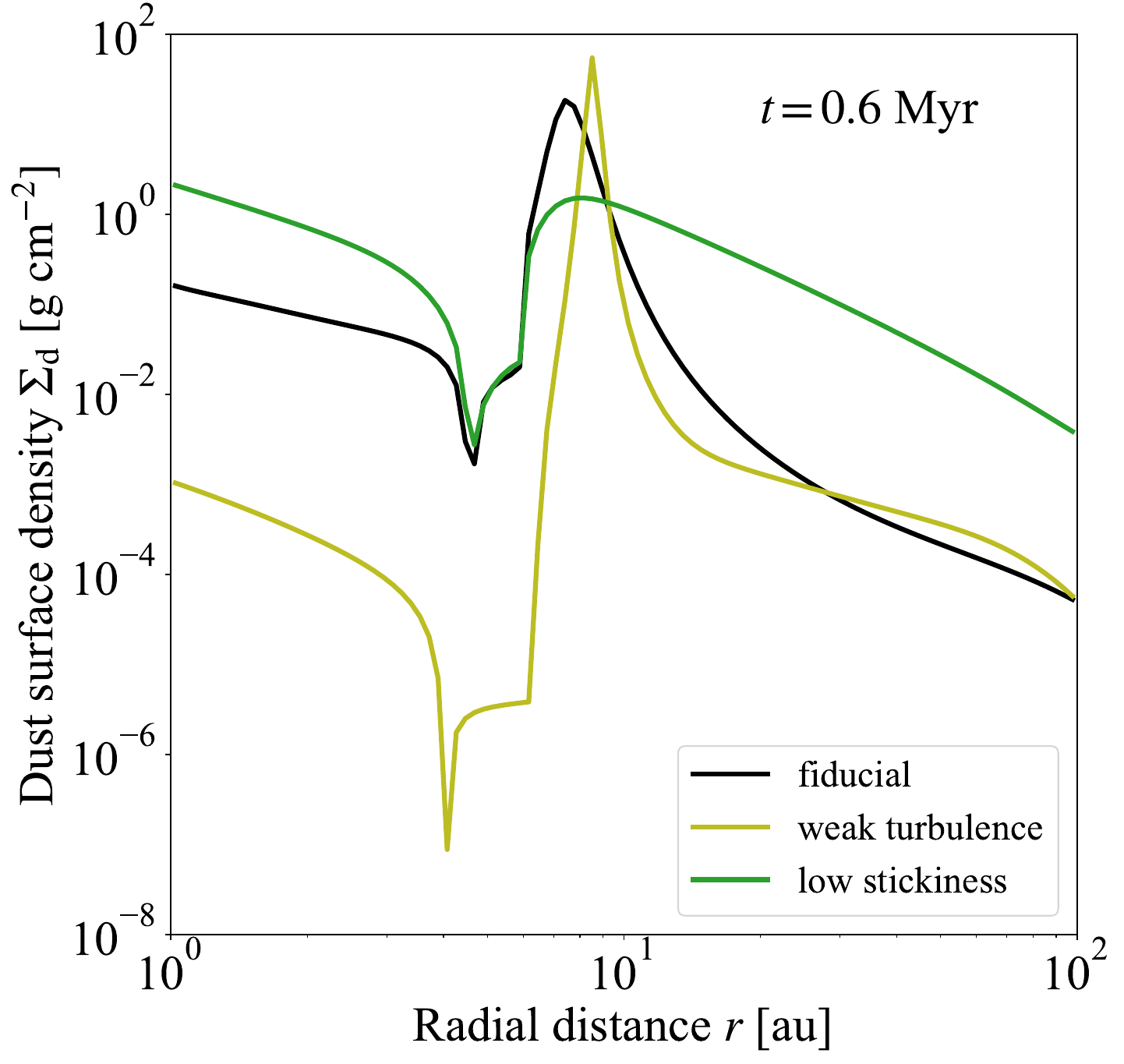}
\includegraphics[width=0.4\hsize, bb = 0 0 691 654]{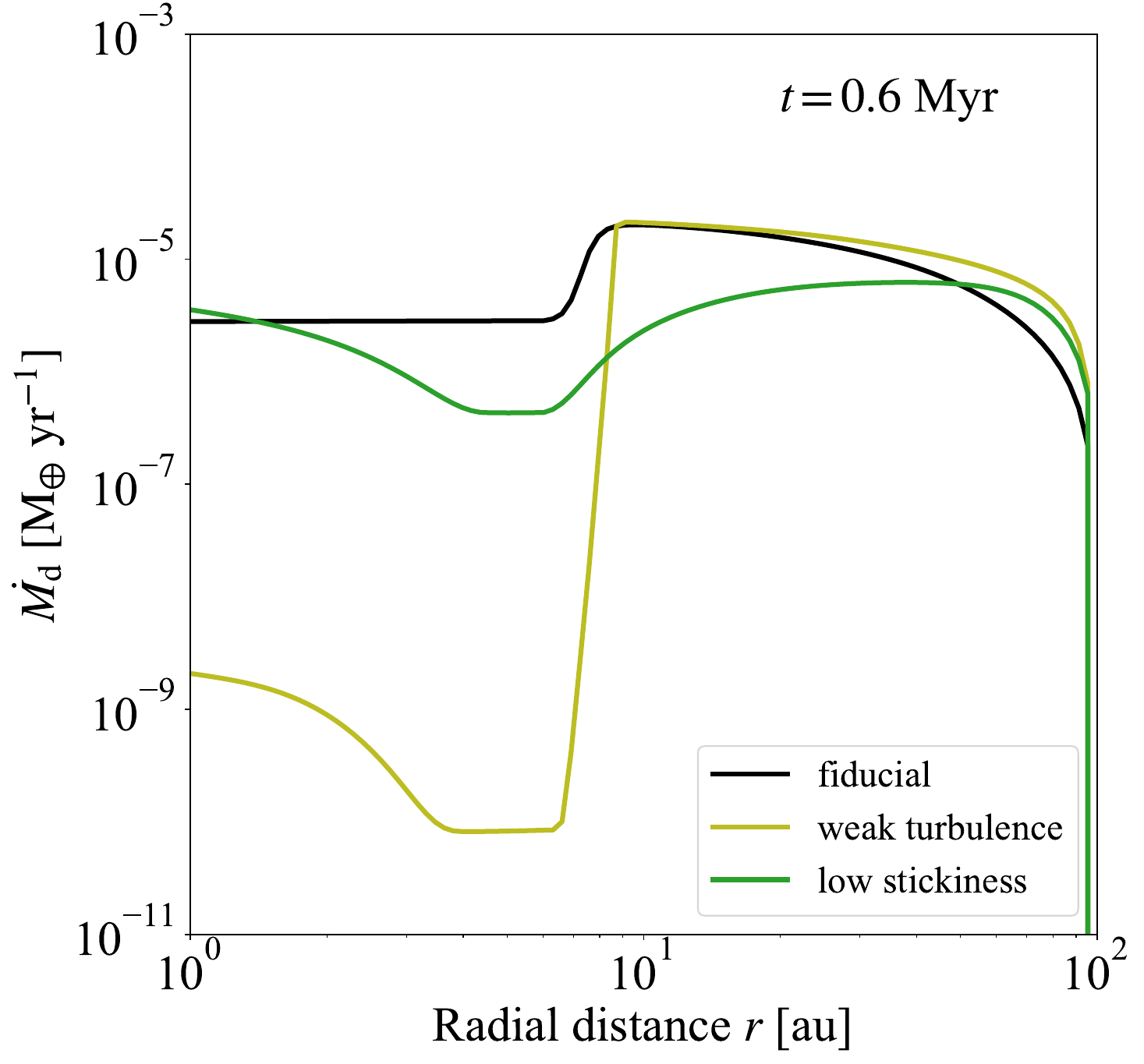}
\end{center}
\caption{Radial distributions of the dust surface density $\Sigma_{\rm d}$ and the size-integrated dust mass accretion rate $\dot{M_{\rm d}}$ at $t = 0.6$ Myr from different simulation runs. Black, yellow, and green lines are from the fiducial run, the weak turbulence run, and the low stickiness run, respectively. 
}\label{f:sigmad}
\end{figure*}

\begin{figure}[ht]
\begin{center}
\includegraphics[width=\hsize, bb = 0 0 750 655]{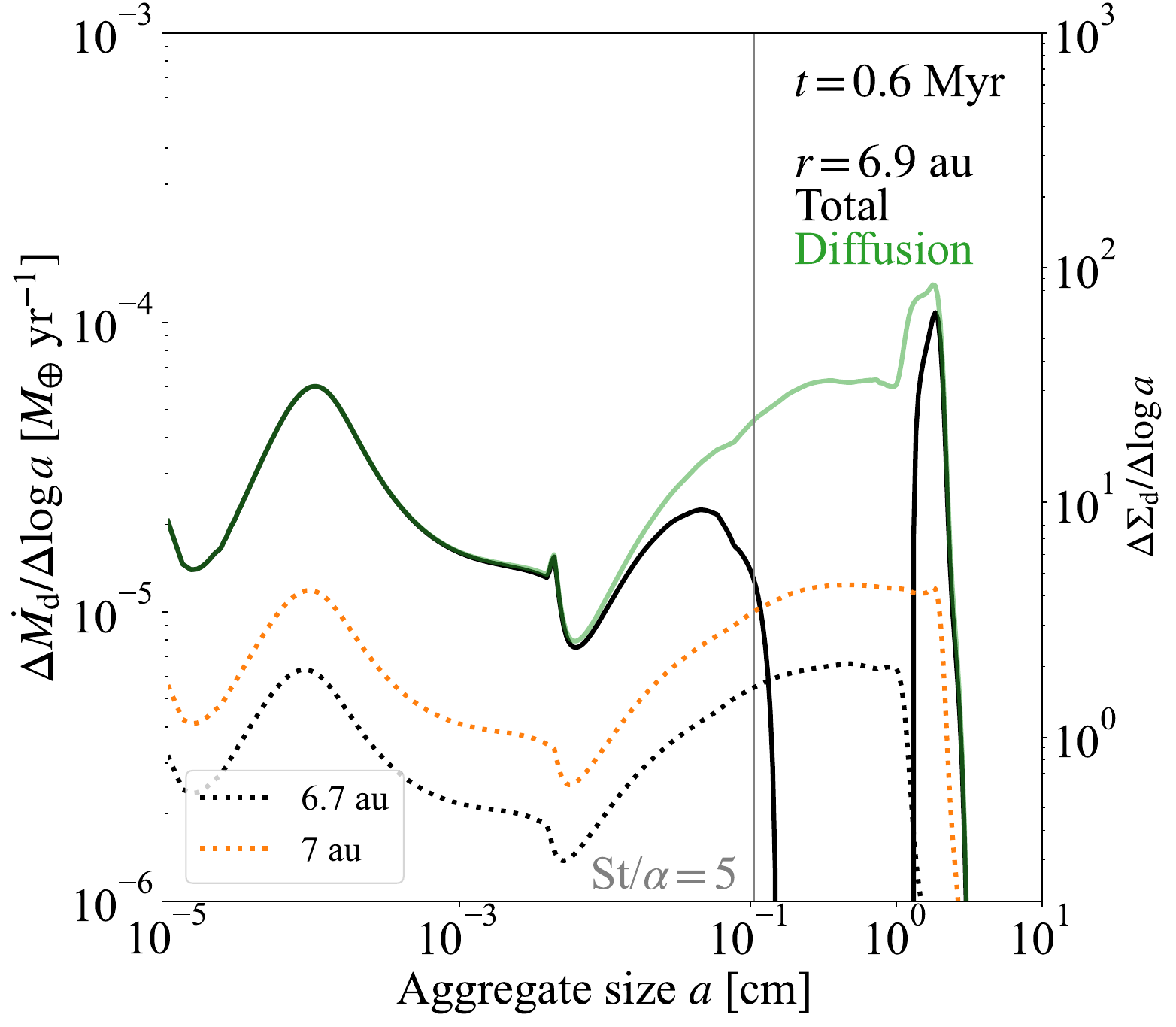}
\end{center}
\caption{
Radial inward dust mass flux per logarithmic dust size, $\Delta \dot{M_{\rm d}}/\Delta \log a$, at the gap outer edge ($r = 6.9$ au) in the fiducial run.
The black line shows the net $\Delta \dot{M_{\rm d}}/\Delta \log a$ due to radial drift and turbulent diffusion, whereas the green line is the contribution from diffusion only. The size corresponding to  ${\rm St}/\alpha =~5$ is shown in the gray horizontal line. 
The black and orange dotted lines show the dust surface density per logarithmic aggregate size, $\Delta \Sigma_{d} / \Delta \log a$, in the fiducial run for $t = 0.6$ Myr at $r = 6.7$ and 7 au, respectively. 
}\label{f:dMddot}
\end{figure}

\begin{figure}[ht]
\begin{center}
\includegraphics[width=\hsize, bb = 0 0 720 720]{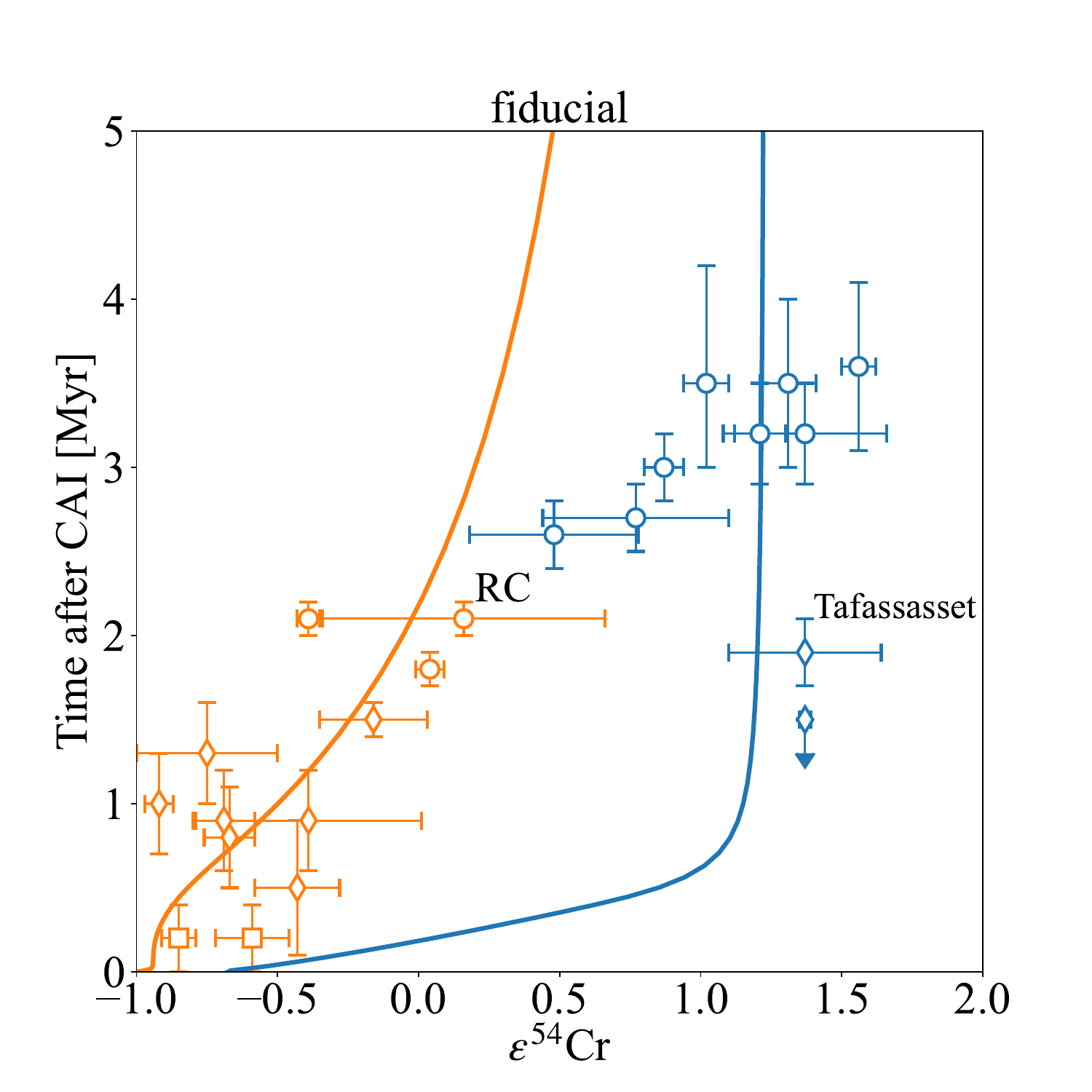}
\end{center}
\caption{Time evolution of $\varepsilon^{54}$Cr values at the NC and the CC reservoirs (orange and blue lines, respectively) in the fiducial run. The orange and blue points indicate the meteoritic data shown in table \ref{t:data}.}\label{f:54Crtime_vf10a3}
\end{figure}

Figure \ref{f:sigmad} displays the radial distribution of $\Sigma_{\rm d}$ at $t = 0.6$ Myr for the weak turbulence and low stickiness runs, respectively.
In comparison with the fiducial run, dust is depleted in the region between $r \approx 1$ and $5$ au in the weak turbulence run. This depletion occurs because the pressure bump at the outer edge of the planet-induced gap traps aggregates completely in the weak turbulence run, unlike in the fiducial run. In the low stickiness run, almost all aggregates leak from the planet-induced gas because the poor-sticky aggregates do not grow to the size that piles up at the pressure bump (see section \ref{s:results_3d} for the conditions of dust accumulation). As a result, the accretion of aggregates in the region exterior to the planet is not observed in the low stickiness run.

\begin{figure}[ht]
\begin{center}
\includegraphics[width=\hsize, bb = 0 0 720 648]{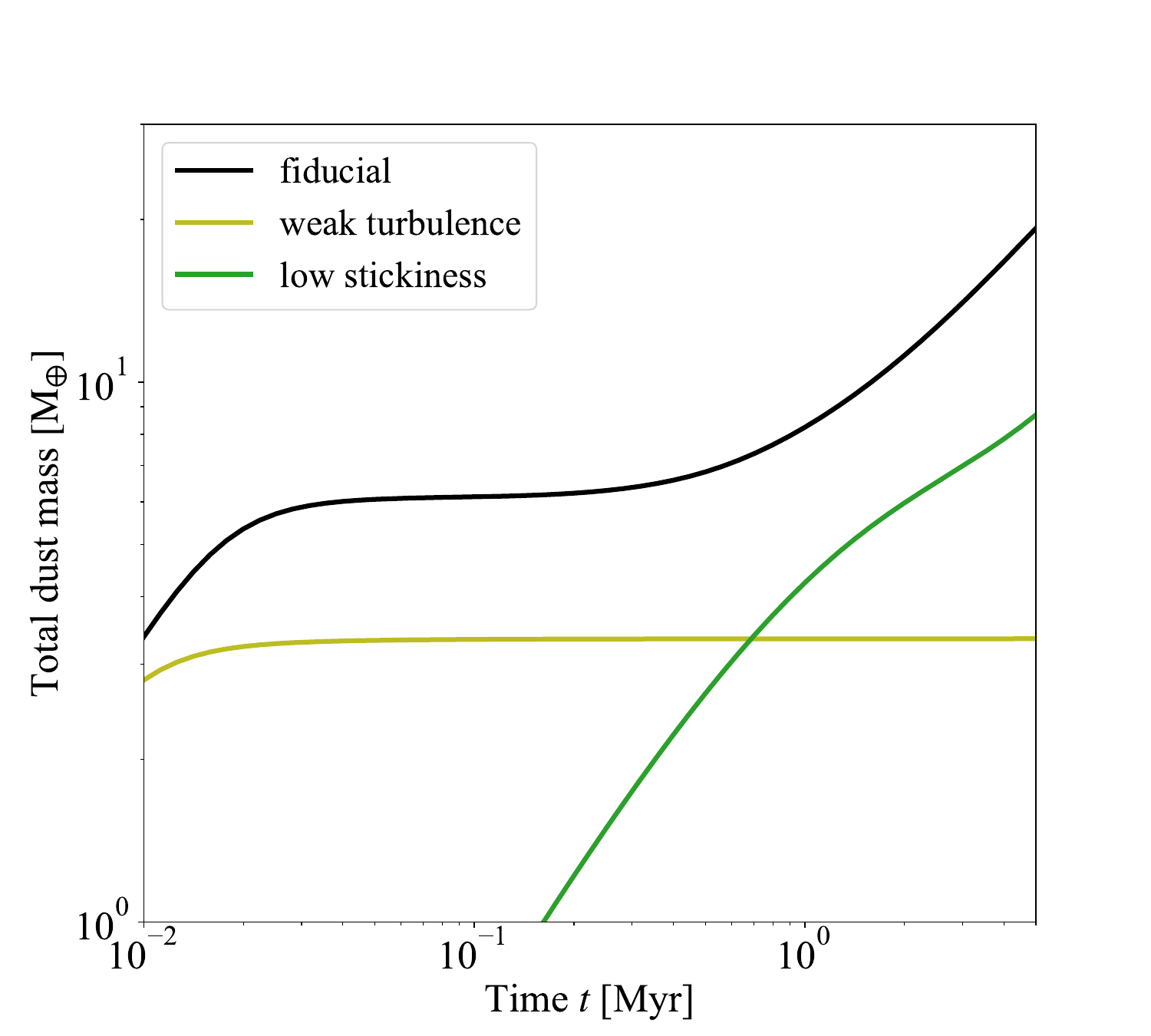}
 \end{center}
\caption{Evolution of total dust masses at the NC reservoir from different simulation runs. The black, yellow, and green lines are the dust masses in the fiducial, weak turbulence, and low stickiness runs, respectively.}\label{f:Md_NC}
\end{figure}

The yellow and green lines in figure \ref{f:Md_NC} show the time evolution of total dust mass at the NC reservoir in the weak turbulence and low stickiness runs, respectively.
In the weak turbulence run, this mass never increases beyond $\approx$ 3 $\rm M_{\oplus} $, which is the initial total dust mass in the region interior to the gap. 
This confirms that the leaking of the dust from the CC reservoir is negligible in the weak turbulence run.

\begin{figure}[ht]
\begin{center}
\includegraphics[width=\hsize, bb = 0 0 720 720]{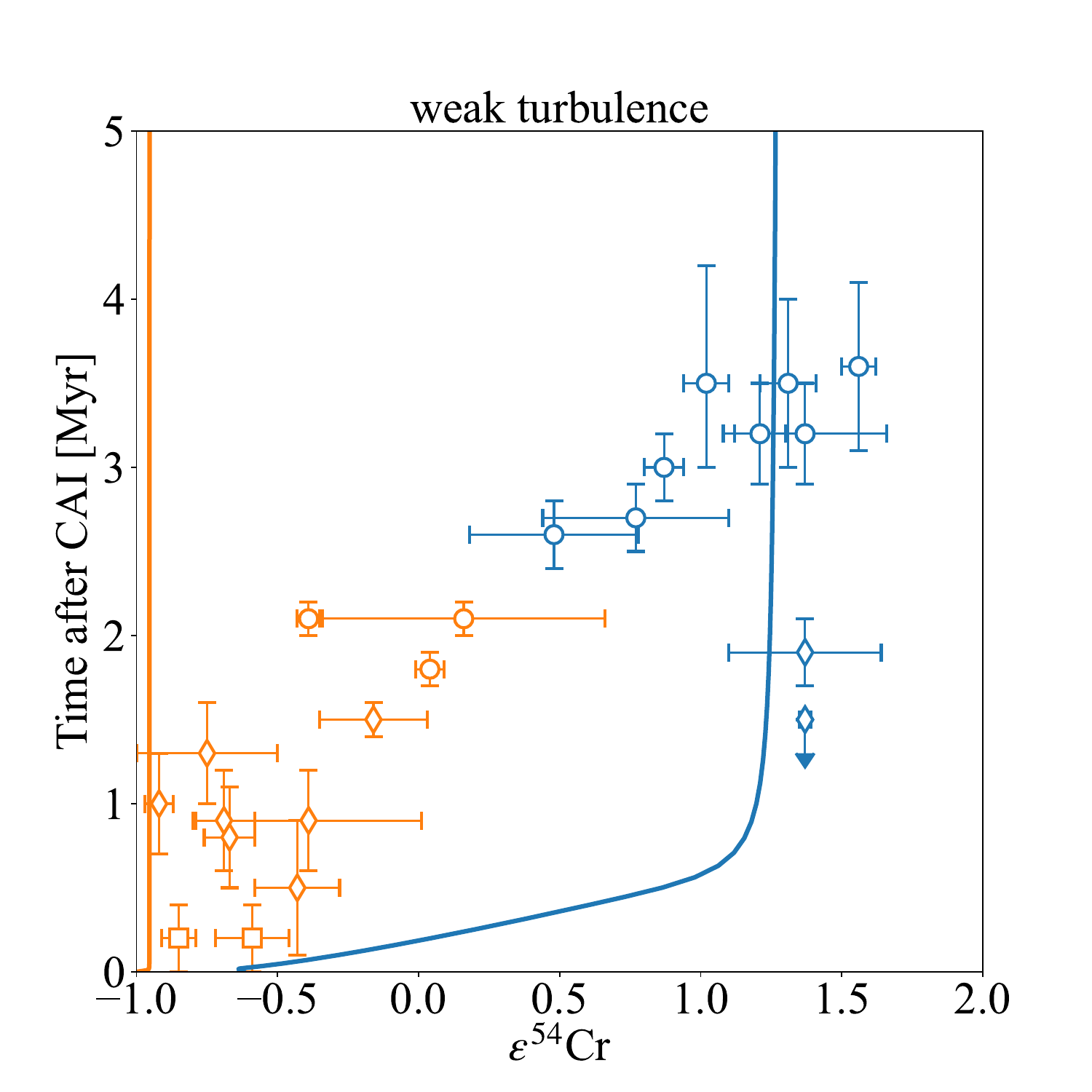}
\end{center}
\caption{Same figures as figure \ref{f:54Crtime_vf10a3} but in the weak turbulence run.}
\label{f:54Cr_vf10a4}
\end{figure}

Figure \ref{f:54Cr_vf10a4} shows the time evolution of the $\varepsilon^{54}$Cr values at the NC and CC reservoirs in the weak turbulence run.
In contrast to the fiducial run, the $\varepsilon^{54}$Cr at the NC reservoir does not increase with time in the weak turbulence run.
This is because the $^{54}$Cr-rich particles at the CC reservoir are allowed little to pass through the gap by diffusion. 
As in the fiducial run, the $\varepsilon^{54}$Cr value at the CC reservoir increases over time because $^{54}$Cr-rich aggregates from the outer region pile up at the gap's outer edge.
The fiducial and weak turbulence runs adopt the same value of $v_{\rm f}$ and therefore show a similar evolution of $\varepsilon^{54}$Cr at the CC reservoir.

We do not show the time evolution of the $\varepsilon^{54}$Cr values in the low stickiness run. 
As shown above, the CC reservoir does not form because aggregates are not accumulated at the bump.

\section{Discussion}\label{s:dis}
\subsection{The solar system isotopic dichotomy}\label{s:d_dich}
Based on the results presented in the previous section, we here discuss whether the EJ scenario can explain the isotopic dichotomy of the solar system. We particularly focus on two key features of the $^{54}$Cr isotopic heterogeneity of meteorites as a function of accretion age (see figure \ref{f:54Crtime_data}):

\begin{enumerate}
\item At accretion ages of $\sim 2$ Myr after CAI formation, there are both NC and CC meteorites (OC, RC, and EC versus Taffassaset and NWA 011) with distinct values of $\varepsilon^{54}$Cr.
\item The $\varepsilon^{54}$Cr values of NC meteorites are positively correlated with accretion age \citep{Sugiura&Fujiya14}. In contrast, such a correlation is less clear for CC meteorites, with Taffassaset and NWA 011 and the other younger samples having similar $\varepsilon^{54}$Cr $\sim +1$...1.5.
\end{enumerate}

The first feature suggests that there was a spatial heterogeneity of $\varepsilon^{54}$Cr in the solar nebula at least at $t \sim 2~\rm Myr$.  
Our simulations confirm that early-formed Jupiter was indeed able to create and sustain the spatial isotopic heterogeneity between NC and CC reservoirs, although some amount of CC materials may have leaked into the NC reservoir depending on turbulence strength and dust stickiness.
In particular, our fiducial run (figure \ref{f:54Crtime_vf10a3}) reproduces $\varepsilon^{54}{\rm Cr} \approx -0.06$ and $\approx 1.20$ for NC and CC reservoirs at $t = $ 2 Myr, consistent with the values of RC and Taffassaset to within their 95\% confidence intervals, respectively. 
We note that the $\varepsilon^{54}$Cr values of the NC and CC reservoirs from our simulations depend on the assumed initial radial distribution of $\varepsilon^{54}$Cr in the solar nebula. 
Our fiducial run demonstrates that there still exists a simple initial radial distribution of $\varepsilon^{54}$Cr (figure \ref{f:54Cr_ini}) that can crudely reproduce the $\varepsilon^{54}$Cr values at the NC and CC reservoirs. 

The second feature can be interpreted as the consequence of the dust leaking through the Jovian gap.  
Our fiducial run indeed reproduces the increase of the NC reservoir's $\varepsilon^{54}$Cr with time thanks to the progressive contamination of dust from the CC reservoir (figure \ref{f:54Crtime_vf10a3}). 
The weak turbulence and the low stickiness runs do not reproduce this trend because the leakage is strongly inhibited and promoted, respectively (section \ref{s:results_FaW}).
These results imply that the correlation between the NC's $\varepsilon^{54}$Cr and accretion age can be used to constrain the range of turbulence strength ($\alpha$) and the stickiness of aggregates ($v_{\rm f}$) in the solar nebula under the assumption that a gas gap was the origin of the solar system isotopic dichotomy.

Although the correlation between $\varepsilon^{54}$Cr and accretion age is weaker for CC meteorites, the relatively low value of $\varepsilon^{54}$Cr of CK chondrites may indicate that the $\varepsilon^{54}$Cr values of CC chondrites (i.e., CC meteorites except Tafassasset and NWA
011) are correlated with their accretion ages as those of NC meteorites \citep{Sugiura&Fujiya14}. Given the large uncertainty of CK chondrites' $\varepsilon^{54}$Cr, it is difficult to ascertain that this correlation is real. Our fiducial simulation (figure \ref{f:54Crtime_vf10a3}) rather supports the interpretation that the  $\varepsilon^{54}$Cr values of CC meteorites, including CO chondrites, Tafassasset, and NWA 011, are more or less independent of time. 

No matter whether the aforementioned time correlation is real or not, it is true that CC meteorites exhibit a variation in $\varepsilon^{54}$Cr that cannot be explained by uncertainties.
This variation is not reproduced in our simulations and may imply that the parental bodies of CC meteorites formed at more than one location in the solar nebula \citep{Fukai&Arakawa21}.
For example, it is possible that the dust trap at Jupiter's gap promoted the formation of Saturn \citep{Kobayashi+12}, which would in turn have created another gas gap and dust trap. 
There are also a number of hydrodynamical instabilities that can produce multiple pressure bumps \citep[e.g.,][]{Bae+23}. 
The parent bodies of CC meteorites of different $\varepsilon^{54}$Cr values may have formed in different dust traps. This possibility should be explored in future work. 

\subsection{Terrestrial planets formation}
The correlation between $\varepsilon^{54}$Cr values and the ages gives a clue about the formation of Earth and Mars. 
By the definition of $\varepsilon^{54}$Cr, the bulk silicate Earth has a $\varepsilon^{54}$Cr value of 0.
Martian meteorites indicate the $\varepsilon^{54}$Cr value is -0.17 $\pm$ 0.03 for the bulk silicate Mars \citep{Trinquier+07}.
By assuming the Earth and Mars are formed at the NC reservoir, our results reveal that the $\varepsilon^{54}$Cr values are $\approx$ -0.17 at 3.2 Myr and $\approx$ 0 at 4.8 Myr at the NC reservoirs in the fiducial run.
We can not estimate the accretion age of Earth and Mars in the weak turbulence run because the $\varepsilon^{54}$Cr value at the NC reservoir is lower than -0.5 for this run.
This suggests that the Earth and Mars accreted the materials at the CC reservoir to reproduce the $\varepsilon^{54}$Cr value.

Our results suggest the rapid formation of Earth and Mars. 
The Hf–W decay system shows that Mars accreted half of its mass within 3 Myr after CAI-formation \citep{Dauphas&Pourmand11}, which aligns with our result for the fiducial run. 
The Hf–W decay system also suggests that the core formation age of Earth is longer than 30 Myr after CAI formation \citep{Kleine+09}.
However, the core formation age of Earth could be reset by the Moon-forming impact \citep{Yu&Jacobsen11, Fischer&Nimmo18, Johansen+23b}.
Iron isotope suggests the rapid formation of proto-Earth in the gas disks \citep{Schiller+20,Johansen+21}.
The accretion age of Earth has still been discussed.
Nonetheless, our results suggest that the rapid accretion of proto-Earth is suitable to reproduce $\varepsilon^{54}$Cr values of the bulk silicate Earth. 

\subsection{$v_{\rm f}$ and $\alpha$}
Our results show that the combination of $v_{\rm f}$ and $\alpha$ determines the radial inward flux of the dust leaking across the Jovian gap.
It is therefore important to discuss whether our adopted values of $v_{\rm f}$ and $\alpha$ are reasonable for the solar nebula.

Recent theoretical studies have indicated that protoplanetary disks are weakly turbulent \citep{Lesur+23}.
This is primarily due to the suppression of disk's instabilities that would drive turbulence. 
For example, MRI remains inactive at the midplane of the disks, except in the inner most regions where $T \gtrsim 1000~\rm K$, due to low ionization rates \citep{Gammie96,Bai&Stone13b}.
The vertical-shear instability, which is a purely hydrodynamic disk instability that operates in disk regions with short cooling timescales  \citep{Nelson+13,Lin&Youdin15}, is suppressed by dust growth because larger dust aggregates are more inefficient at cooling the disk gas \citep{Malygin+17,Fukuhara+21}.
Hydrodynamic simulations suggest that the value of $\alpha$ in such weakly turbulent disks would likely be lower than $10^{-3}$ \citep{Lesur+23}, which we adopted in the fiducial run.
As shown in the weak turbulence run, aggregates with $ v_{\rm f} = 10 \rm~m~s^{-1}$ accumulate at the bump without moderate dust leaking required to reproduce the temporal increase of the NC reservoir’s $\varepsilon^{54}$Cr if $\rm \alpha \ll 10^{-3}$.

Even in weakly turbulent disks, the moderate leaking could occur if the aggregates are poorly sticky.
The stickiness of aggregates is determined by the materials on the surface of monomers \citep{Dominik&Tielens97,Wada+13}.
In the outer and cool part of the disk, $\rm H_{2}O$ and/or $\rm CO_{2}$ freeze on the surface of monomers.
We assumed $v_{\rm f} = 10 \rm ~m~s^{-1}$, which might be about one order of magnitude higher than the value of $v_{\rm f}$ estimated by the previous laboratory experiments with these ice-mantled particles \citep{Musiolik+16, Fritscher&Teiser21}.
Moreover, the no-sticky ice is favorable for explaining the polarimetric observation of the dust disk around HL Tau \citep{Okuzumi&Tazaki19}. 
If aggregates are composed of such non-sticky ice-manteled grains, the aggregates experience frequent fragmentation at $\alpha \ll 10^{3}$.
Therefore, a moderate amount of the poor-sticky dust particles could leak from the CC reservoir.
This leakage potentially explains the observed correlation of the NC reservoir in a weakly turbulent disk.

\subsection{Processes not included in the present model}\label{s:d_pro}
This work is our first step toward fully understanding the roles of dust collisional evolution in the isotopic compositional gradients in protoplanetary disks. For this reason, we have ignored a number of gas and dust processes that potentially affect the mass and isotopic evolution of the NC and CC reservoirs. 
In this section, we discuss the potential significance of two neglected processes: planetesimal formation and dust leaking from the NC reservoir.

\subsubsection{Planetesimal formation}
\begin{figure}
    \begin{center}
    \includegraphics[width=\hsize,bb = 0 0 684 248]{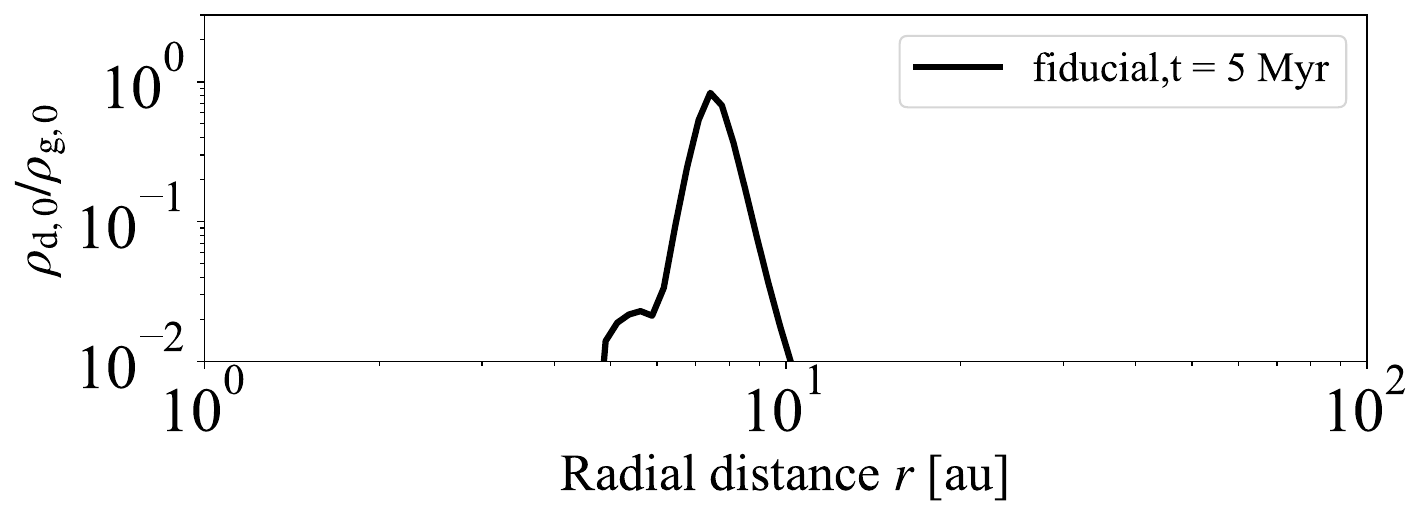}
    \end{center}
    \caption{Ratio of the size-integrated dust density to the gas density at the midplane $\rho_{\rm d,0}/\rho_{\rm g,0}$ in the fiducial run at $t = 5 $ Myr. The maximum value of $\rho_{\rm d,0}/\rho_{\rm g,0}$ is almost constant over $t_{\rm drift}$ at 100 au $\approx 0.3$ Myr.}
    \label{f:dtg}
\end{figure}

In general, local accumulation of solids can lead to planetesimal formation via the streaming instability \citep{Youdin&Goodman05,Johansen+07,Yang+17} and/or the gravitational instability \citep{Goldreich&Ward73, Sekiya83}.
In our simulations, the outer edge of the planet-induced gap traps large aggregates and may thus convert them into planetesimals.
In the fiducial run, the size-integrated dust density at the midplane at the outer edge of the planet-induced gap reaches the midplane gas density (figure \ref{f:dtg}).
For the largest aggregates with ${\rm St} \approx 0.1$, the midplane dust-to-gas mass ratio of $\approx 1$ is barely enough to trigger strong dust clumping by the SI  (see figure 4 of \citealt{Li+21}). 

Planetesimal formation at the gap's outer edge (the CC reservoir in our model) naturally explains the formation of solid bodies with a CC-like isotopic composition. On the other hand, this process would also affect the isotopic compositional evolution of the CC reservoir because the mass balance between the remaining solids and the solids newly added to the reservoir through radial inward drift would change. Because solids that arrive at the gap edge at later times have a higher $\varepsilon^{54}$Cr value, planetesimal formation would result in a further increase in the $\varepsilon^{54}$Cr of the remaining solids with time. Planetesimals that form earlier would have a lower $\varepsilon^{54}$Cr value. This would also be true for planetesimals forming in the NC reservoir if the small particles are allowed to leak from outside to inside the gas gap as in our fiducial run.  

\subsubsection{Gas velocity} \label{s:vg}
In our simulations, turbulent diffusion leaks particles from the pressure bump outside the planet-induced gas gap by turbulent diffusion. 
However, particles could leak not only by diffusion but also by the accretion of the disk.
Here, we compare the diffusion and accretion timescales to show that the latter is indeed negligible.

Assuming that disk accretion is in a steady state, the gas accretion velocity $v_{\rm g}$ can be rewritten as $v_{\rm g} = -3 \nu /(2r)$ \citep{Lynden-Bell&Pringle74}, where $\nu$ is the viscosity of the disk. 
If ${\rm St} \ll 1$, this velocity adds to the dust radial velocity \citep{Takeuchi&Lin02}.
Because particles at ${\rm St} < 5 \alpha \ll 1$ leak from the bumps, the accretion timescale of the particles $t_{\rm acc}$ is given by $t_{\rm acc} = r/v_{\rm g} \sim r^2/\nu$. 
The diffusion timescale of particles is given by $t_{\rm diff} = r/D_{\rm d} \sim r^2/\nu$.
The ratio of $t_{\rm acc}$ to $t_{\rm diff}$ is estimated as $t_{\rm acc}/t_{\rm diff} \sim \nu/D_{\rm d}$.
Using the alpha model of \cite{shakura&Sunyaev73}, $\nu$ is comparable to $D_{\rm d}$ for particles,  except for the region inside the planetary gap.
Therefore, the ratio of the two timescales would be unity at the bump outside the gap.
This estimation suggests that the amount of dust leaking from the gap by gas accretion is comparable to that by turbulent diffusion.
Therefore, it does not significantly impact our conclusion to ignore gas accretion.

\subsubsection{Dust leaking from the NC reservoir} \label{s:d_inn}
In our simulations, we did not explicitly model the NC reservoir and instead assumed that all dust drifting inward through the inner boundary of the computational domain will eventually be trapped at the the NC reservoir somewhere inward of the computational boundary. 
If a pressure bump is responsible for the dust accumulation at the NC reservoir, the small dust particles may pass through the bump as those at the CC reservoir do in the fiducial run.
If the dust at the NC reservoir were to leak, the $\varepsilon^{54}$Cr value at the reservoir could be higher than that derived from our simulation results.
This is because the reservoir would then lose a larger amount of dust that arrived at earlier times, which are assumed to have lower $^{54}$Cr values in our model. 
By explicitly assuming a potential mechanism for the formation of the NC reservoir, it is necessary to examine whether the $\varepsilon^{54}$Cr value at the NC reservoir is consistent with those of NC meteorites even if dust leaks from there.

Our simulations also prohibit the dust at the NC reservoir from diffusing back to the computational domain.
In reality, there would exist a positive radial gradient of $\varepsilon^{54}$Cr between the $^{54}$Cr-poor NC reservoir and the $^{54}$Cr-richer computational domain, which would produce an outward diffusion flux of $^{54}$Cr particles from the reservoir to the computational domain. 
However, under the assumption that the parent bodies of NC meteorites were formed at the NC reservoir, this would not alter the evolution of $^{54}$Cr for these meteorites.
The $^{54}$Cr abundance of the NC reservoir is dominated by the total amount of the dust particles leaking from the CC reservoir (see figure \ref{f:Md_NC} and section \ref{s:Crevol_vf10a3}).
Therefore, the outward diffusion flux is not likely to alter the value of $\varepsilon^{54}$Cr at the NC reservoir.

\section{Summary}\label{s:summary}
This study explored how dust leaking across the Jovian gap in the solar nebula influences the isotopic ratios of NC and CC meteorites.
We constructed a model to simulate the evolution of dust size distribution and $^{54}$Cr abundance through collisional coagulation and fragmentation in protoplanetary disks (section \ref{s:methods}).
We assumed that the parent bodies of NC and CC meteorites formed at the two distinct locations where dust aggregates pile up interior and exterior to Jupiter, namely the NC reservoir and CC reservoir (section \ref{s:eval} and figure \ref{f:sch}).  We assessed the temporal variation of $\varepsilon^{54}$Cr at these reservoirs due to the dust leaking, considering that the value of $\varepsilon^{54}$Cr is higher toward the outer region of the disk.

Our key findings are summarized as follows:
\begin{enumerate}
\item We found that aggregates at $\rm St/\alpha > 5$ pile at the pressure bump created by the Jovian gap (section \ref{s:results_3d}). This criterion aligns with the estimation based on the balance of radial drift and turbulent diffusion \citep{Zhu+12}.

\item While large aggregates pile up at the bump and the CC reservoir forms there, the NC reservoir accumulates a significant amount of dust leaked from the CC reservoir.
Our fiducial run shows that dust fragments resulting from the collisional fragmentation of aggregates at the bump leak across the gap through turbulent diffusion (section \ref{s:results_3d} and figure \ref{f:Md_NC}).
We also illustrated the dependence of the dust leaking on turbulence strength and dust stickiness (section \ref{s:results_FaW}).

\item  The value of $\varepsilon^{54}$Cr at the CC reservoir evolves until the drift timescale of the aggregates at the outer boundary of disks (see section \ref{s:Crevol_vf10a3}). 
The continuous dust leaking from the CC reservoir affects little the evolution of $\varepsilon^{54}$Cr at the CC reservoir (the blue line in figure \ref{f:54Crtime_vf10a3}). 

\item The continuous leaking increases the values of $\varepsilon^{54}$Cr at the NC reservoir over time (the orange line in figure \ref{f:54Crtime_vf10a3}).
However, our fiducial run reproduces the values of $\varepsilon^{54}$Cr for the NC and CC reservoirs at $t$ = 2 Myr which are consistent with the values of RC and Taffassaset. respectively.
This finding indicates that the early-formed Jupiter is able to create and sustain the spatial isotopic heterogeneity between the NC and CC reservoirs even if the NC reservoir accumulates a substantial amount of dust leaking from the CC reservoir (section \ref{s:d_dich}).
Moreover, this dust leaking reproduces the observed temporal variation of $\varepsilon^{54}$Cr for NC meteorites.

\item Our model does not reproduce the temporal variation of $\varepsilon^{54}$Cr for CC meteorites. This variation may be produced by the formation of the parent bodies of these meteorites in different dust traps.
\end{enumerate}
 
Gas and dust processes that could potentially affect the mass and isotopic evolution at the NC and CC reservoirs are not considered in our model (section \ref{s:d_pro}).
To comprehensively understand these evolutions at the two reservoirs, it is necessary to account for such processes, including planetesimal formation and the leakage of dust from the NC reservoir.
\begin{ack} 
We thank Tetsuya Yokoyama, Shigeru Ida, Taishi Nakamoto, Yuki Masuda, Kanon Nakazawa, and Emmanuel Jacquet for useful comments and discussions about the materials of the solar system and disk evolution. 
We also thank the anonymous referee for comments that significantly improved this paper.
This work was supported by JSPS KAKENHI Grant Numbers JP20H00182, JP20H00205, 20K14535, JP22KJ1310, JP23H00143, JP23H01227, JP23K25923, and JP24K17118. 
\end{ack}
\bibliographystyle{apj}
\bibliography{bib}
\end{document}